**Title:** Dissolution zone model of the oxide structure in additively manufactured dispersion-strengthened alloys

**Authors:** Wenyuan Hou[a], Timothy Stubbs[a,b], Lisa DeBeer-Schmitt[c], Yen-Ting Chang[d], Marie-Agathe Charpagne[d], Timothy M. Smith[e], Aijun Huang[b], Zachary C. Cordero[a]

**Affiliation:**

a. Aeronautics and Astronautics, MIT, Cambridge, MA 02139, USA
b. Monash Centre for Additive Manufacturing, Monash University, Notting Hill, VIC 3168, Australia
c. Neutron Scattering Division, Oak Ridge National Laboratory, Oak Ridge, TN 37831, USA
d. Materials Science and Engineering, University of Illinois, Urbana, IL 61801, USA
e. NASA Glenn Research Center, Cleveland, OH 44135, USA

**Corresponding author:** Zachary Cordero, zcordero@mit.edu, +1 (617) 253-8821

**Abstract:** The structural evolution of oxides in dispersion-strengthened superalloys during laser-powder bed fusion is considered in detail. Alloy chemistry and process parameter effects on oxide structure are assessed through a parameter study on the model alloy Ni-20Cr, doped with varying concentrations of $Y_2O_3$ and Al. A scaling analysis of mass and momentum transport within the melt pool, presented here, establishes that diffusional structural evolution mechanisms dominate for nanoscale dispersoids, while fluid forces and advection become significant for larger micron-scale slag inclusions. These findings are developed into a theory of dispersoid structural evolution, integrating quantitative models of diffusional processes – dispersoid dissolution, nucleation, growth, coarsening – with a reduced order model of time-temperature trajectories of fluid parcels within the melt pool. Calculations of the dispersoid size in single-pass melting reveal a zone in the center of the melt track in which the oxide feedstock fully dissolves. Within this zone the final $Y_2O_3$ size is independent of feedstock size and determined by nucleation and growth kinetics. If the dissolution zones of adjacent melt tracks overlap sufficiently with each other to dissolve large oxides, formed during printing or present in the powder feedstock, then the dispersoid structure throughout the build volume is homogeneous and matches that from a single pass within the dissolution zone. Gaps between adjacent dissolution zones result in oxide accumulation into larger slag inclusions. Predictions of final dispersoid size and slag formation using this dissolution zone model match the present experimental data and explain process-structure linkages speculated in the open literature.

**Keywords:** laser-powder bed fusion; oxide dispersion strengthening; nickel-based superalloys; neutron scattering

## 1. Introduction

Oxide dispersion-strengthened (ODS) alloys contain thermally stable oxide nanoparticles that obstruct dislocation motion at elevated temperatures, enhancing creep resistance [1,2]. Several recent studies [3–7] have attempted to print net-shaped ODS alloys using laser-powder bed fusion (L-PBF), under the premise that the brief melt time will preserve nanoscale dispersoids [8]. However, dispersoids in L-PBF ODS alloys tend to be larger than those in wrought materials [8,9]. In addition, certain ODS alloys form large slag inclusions during printing [7,10]. Mitigating these issues is essential to printing dispersion-strengthened alloys for demanding structural applications.

Experiments have established that two key factors – melt time [6,10,11] and alloy chemistry [7,10,12] – strongly influence the final dispersoid structure in L-PBF ODS materials. Generally, reducing the melt time yields smaller dispersoids [6,10,11]. By optimizing printing parameters, feedstock morphology, and component geometry, dispersoid sizes as small as 20–150 nm have been achieved [5–7,10,13–16]. Alloy chemistry also affects dispersoid structure, since dispersoids can react with alloying additions, forming mixed oxides with distinct coarsening behaviors. Such reactions have important implications for alloy design. For example, Ni-base ODS alloys achieve γ'-strengthening through Al alloying. However, even low concentrations of Al can react with $Y_2O_3$ dispersoids during L-PBF, forming Y-Al-O compounds with accelerated coarsening kinetics [17]. This phenomenon was observed in L-PBF MA754 (Ni-20Cr-0.34Al-0.43Ti-0.6$Y_2O_3$, wt%), where slag inclusions comprised Y-Al-O phases that formed *in situ,* providing direct evidence that Al promotes slag [10]. Similar behaviors have also been observed in ferrous ODS alloys [18,19] and in Ni-base alloys with $HfO_2$ dispersoids [20]. These observations illustrate the need to co-optimize printing parameters and alloy chemistry to achieve nanoscale dispersoids.

The two main proposed mechanisms of dispersoid growth during melt-based additive manufacturing of ODS alloys are: (i) mechanical impingement [6,14] and (ii) chemical transport [6,10,11,14]. In mechanical impingement, forces act on dispersoids, driving them together within the melt pool [6,8,14]. To gauge the significance of these fluid forces, dimensionless groups that describe the flow around the particle are computed below (cf. **Sec. 4.1**) for the ranges of fluid velocities and particle sizes encountered in practice. A key takeaway from this scaling analysis is that dispersoids smaller than 100 nm experience creeping flow, with no relative motion between the particle and nearby fluid. This in turn suggests there is no relative motion between nearby particles, indicating that fluid forces cannot drive mechanical impingement of such nanoscale dispersoids. Another potential source of mechanical impingement is dispersoid pushing by the solidification front. However, the fast solidification velocity in L-PBF traps even nano-scale dispersoids [21], suggesting particle pushing via solidification is also insignificant in this process. This prediction aligns with observations of a uniform spatial distribution of dispersoids in the as-printed material, with no apparent dispersoid clustering in interdendritic regions. Taken together, these points indicate that mechanical impingement within the melt pool is unlikely to account for coarsening of dispersoids smaller than 100 nm, although it may play an important role in the agglomeration of larger dispersoids into slag inclusions.

An alternative physical description of dispersoid growth is with chemical transport-driven phenomena such as dispersoid dissolution, nucleation, coarsening, and growth [6,14]. These processes are extremely temperature-sensitive and are therefore influenced by the complex time-temperature profile experienced by dispersoids as they traverse the melt pool. The precise thermal excursion experienced by each dispersoid depends on its initial position as it enters the melt. Typically, the fluid temperature rises rapidly as the material passes under the laser, followed by more gradual cooling in the tail of the melt pool. Each phase of this thermal cycle is dominated by different transport processes. During the heating and initial cooling stages, dispersoids dissolve when the oxide solubility exceeds the oxide concentration. The extent of dissolution depends on initial dispersoid size, and despite the brief melt cycle in L-PBF, nanoscale dispersoids can fully dissolve [6]. Subsequently, as the melt cools, dispersoids nucleate from the supersaturated solution, then coarsen and grow. Unlike the mechanical impingement theory, these transport-driven phenomena depend on chemistry-dependent material properties (e.g., solubility [22], interfacial energy [23], melting point, among others) and can therefore account for experimental observations of chemical effects on dispersoid evolution. Further, complete dissolution followed by nucleation and growth can explain why dispersoids sometimes appear to shrink during printing [6,13,19].

Modeling transport-driven dispersoid evolution requires integrating descriptions of fluid flow and heat transfer within the melt pool with chemistry-sensitive structural evolution models. In an early demonstration of such a multi-physics modeling approach, Hong et al. [22] described the formation and growth of oxide inclusions during single-track arc-welding of ferrous alloys. They integrated structural evolution models with computational fluid dynamics simulations that provided the time-temperature profile of individual fluid elements within the melt. Although Hong et al. considered arc welding, several of their findings align with behaviors seen in L-PBF. For example, they found that the time-temperature profiles experienced by each dispersoid varies strongly depending on its initial position and that longer melt times result in larger inclusions.

Similar approaches have recently been used to model dispersoid evolution during L-PBF of ODS alloys. Eo et al. [14] used aspects of Hong et al.'s framework to describe dispersoid formation during reactive gas L-PBF of ferrous alloys. While their analysis accurately predicted dispersoid size for a single set of L-PBF processing conditions, it was subject to several key limitations. First, the model did not account for convection within the melt pool, which tends to contract the thermal excursion experienced by dispersoids moving along streamlines [24]. Second, it only considered a single representative time-temperature profile. Third, the analysis only considered the cooling phase of the melt cycle. These latter two limitations prevent analysis of larger dispersoids capable of surviving dissolution during a single melt pass and subsequently coarsening over multiple passes to form slag.

In a related study, Wassermann et al. [6] modeled $Y_2O_3$ dispersoid evolution during L-PBF of Ni-base ODS alloys. They used thermographic measurements of a single time-temperature profile on the surface of the melt pool as inputs for numerical models of dispersoid evolution. Their analysis showed that process conditions which minimized the melt time (e.g., lower laser power, faster scan speed) yielded smaller dispersoids, consistent with their experimental results. However, their approach faced several limitations: (i) it lacked a description of fluid motion and heat transfer within the melt pool, (ii) it relied on a single time-temperature profile, and (iii) it considered only a single initial dispersoid size. Additionally, the specific combinations of thermal excursion and dispersoid size considered in their study led to complete dispersoid dissolution, rendering their approach unable to predict slag formation for the same reasons observed in the Eo et al. model. Further, the lack of a quantitative description of the melt behavior complicates generalization to a range of processing conditions beyond those explicitly tested.

The discussion above highlights how a complete physical description of dispersoid evolution during L-PBF must consider: (i) time-temperature trajectories experienced by individual fluid elements as they traverse the melt; (ii) thermochemistry-related effects such as reactions between alloying elements and dispersoids; and (iii) cyclic dissolution and growth in regions subjected to multiple melt cycles. In the present work we develop and experimentally validate a modeling framework that addresses these requirements. We first conduct an L-PBF parameter study using a model alloy, Ni-20Cr, doped with varying concentrations of $Y_2O_3$ and Al, to assess the effects of material chemistry and print parameters on final dispersoid size. We use small angle neutron scattering (SANS) to characterize the dispersoid size distribution in the as-printed materials. SANS can probe macroscale volumes, thus providing statistically accurate measurements of the entire dispersoid size distribution.

These experimental measurements serve as the basis for developing a quantitative model of dispersoid evolution during L-PBF, accounting for the essential physics mentioned above. We use computational fluid dynamics simulations to calibrate a reduced order model of location-specific time-temperature trajectories within the melt pool. We then couple this model to chemistry-sensitive structural evolution models of dispersoid dissolution, nucleation, growth, and coarsening. The resulting framework can predict whether dispersoids fully dissolve based on the combination of initial size, initial location, alloy chemistry, and printing parameters. For parameter combinations that result in full dissolution, the framework predicts the final dispersoid size after precipitation and cooling in the tail of the melt pool. Conversely, when the dispersoids do not dissolve, the framework indicates when the dispersoids may agglomerate into larger slag inclusions. These results are synthesized into process diagrams which can be used for alloy design and parameter selection, with a view towards printing fully dense, slag-free materials featuring uniformly distributed nanoscale dispersoids.

## 2. Materials and methods

### 2.1. *Powder feedstock*

ODS powder feedstocks were prepared by decorating gas-atomized Ni-20Cr, wt% powder with $Y_2O_3$ nanopowder via resonant acoustic mixing (RAM), following the procedure in [4]. **Fig. 1** shows Ni-20Cr powder coated with $Y_2O_3$ dispersoids. The high-magnification inset shows larger micron-scale $Y_2O_3$ clusters as well as nanoparticles, likely formed through comminution of larger clusters during RAM. The Ni-20Cr powder had a size distribution between 10 and 54 μm, with a mean particle size of 20 μm. RAM-coating with $Y_2O_3$ preserved the size distribution, spheroidal shape, and flowability of the powder feedstock. Al alloying was accomplished by mechanically blending gas-atomized equiatomic NiAl powder with the ODS feedstock. The different powder compositions are summarized in **Table 1**.

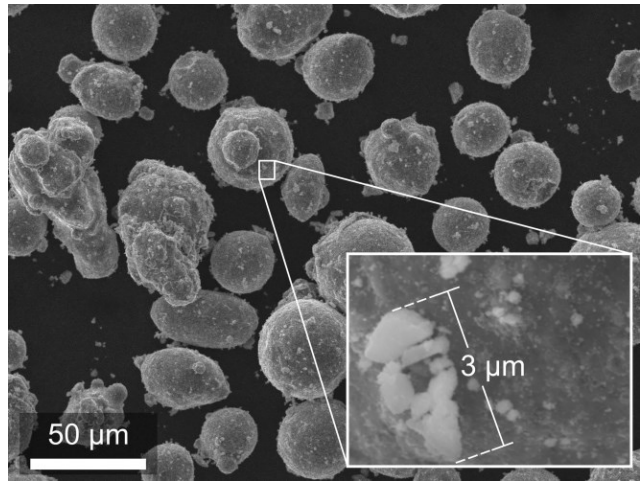

**Fig. 1.** SEM micrograph of RAM Ni-20Cr-0.45Y$_2$O$_3$ feedstock, with inset showing micron-scale Y$_2$O$_3$ agglomerates as well as nanoscale Y$_2$O$_3$ dispersoids.

**Table 1.** Powder compositions.

| Alloy Designation | Composition (wt. %) | | | | | | |
|---|---|---|---|---|---|---|---|
| | Ni | Cr | Al | Y$_2$O$_3$ | Fe | Si | C |
| Ni-20Cr | bal. | 20.2 | - | - | 0.08 | 0.05 | 0.02 |
| Ni-20Cr-1Y$_2$O$_3$ | | | | +1.0 | | | |
| Ni-20Cr-0.45Y$_2$O$_3$ | | | | +0.45 | | | |
| Ni-20Cr-0.45Y$_2$O$_3$-0.34Al | | | +0.34 | +0.45 | | | |
| Ni-20Cr-0.45Y$_2$O$_3$-1Al | | | +1.0 | +0.45 | | | |

## 2.2. *Printing parameters*

Test cubes (1x1x1 cm$^3$) were printed using an EOS M100 L-PBF printer equipped with a ~1040 nm Yb fiber laser (beam power: 200 W; beam radius: 20 μm) in an inert Ar atmosphere (~0.1% O$_2$). All samples were produced with a 5 mm striped scanning strategy, 67° interlayer rotation, 40 μm hatch spacing, and 20 μm layer thickness. Parameter studies were conducted in which the laser power and scan speed were independently varied over the ranges 0.6–2.2 m/s and 60–170 W.

## 2.3. *Structural characterization*

The relative density of the test cubes was assessed using Archimedes' method as well as stereological techniques. Select specimens were mounted, cross-sectioned, polished, and imaged using optical and electron microscopy. Optical microscopy was used to assess lack of fusion defects, porosity, cracks, and the presence of slag inclusions. Scanning electron microscopy (SEM) was used to characterize larger oxide dispersoids as well as slag inclusions. Wavelength dispersive spectroscopy (WDS) measurements were used to compare the composition of the as-printed material with the composition of the feedstock.

Thin lamellar specimens were extracted from the as-printed material, ground, then thinned via electropolishing and imaged in a Talos F200X G2 scanning transmission electron microscope (STEM) equipped with a Super-X EDS system. These STEM measurements were used to characterize the shape, size distribution, and composition of nanoscale dispersoids.

*2.4. Small angle neutron scattering (SANS)*

SANS was used to characterize the dispersoid structure in the as-printed specimens. The samples were analyzed on the HFIR CG2 (GP-SANS) beamline at Oak Ridge National Laboratory. Each scattering spectra was collected over two hours using standard configurations: collimation 17m / SDD 19m / 12Å; collimation 6m / SDD 6m / 4.75Å; and collimation 6m / SDD 1m / 4.75Å. The scattering spectra were analyzed using IRENA [25] to determine the dispersoid size distributions, assuming spherical dispersoids, and the dispersoid volume fraction, assuming the scattering length density contrast between $Y_2O_3$ and Ni-20Cr was $|\Delta\rho|^2 = 1.2 \times 10^{21}$ cm$^{-4}$ [26].

3. **Experimental results and discussion**

*3.1. L-PBF parameter studies*

Viable printing conditions for the Ni-20Cr and Ni-20Cr-1$Y_2O_3$ materials were determined through parameter studies in which the beam power and scan speed were independently varied. **Fig. 2** summarizes the relative density ($\tilde{\rho}$) measurements from these parameter studies. The baseline Ni-20Cr material achieved full density over a wide processing window – scan speeds between 0.9 and 1.8 m/s and laser powers between 90 and 160 W. The ODS material had a narrower processing window (0.8–1.2 m/s; 120–160 W), requiring slower scan speeds and higher powers to achieve full density. This result is consistent with the findings in [15,27] where successful consolidation of RAM ODS materials required higher energy densities than their non-ODS counterparts. This trend was linked to a reduction in laser absorptivity from the oxide nanoparticles on the powder feedstock surface [8,28]. Parameters within the Ni-20Cr-1$Y_2O_3$ processing window (0.8–1.6 m/s; 100–170 W) were used to print materials with varying Al and $Y_2O_3$ concentrations. Modifying the composition did not affect printability or degrade density; however, it did affect the dispersoid size distribution as discussed below.

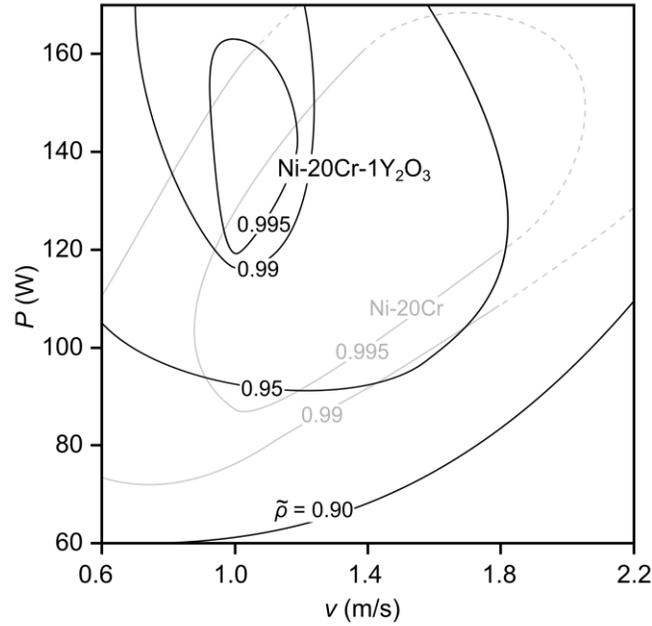

**Fig. 2.** Isocontours of relative density ($\tilde{\rho}$) from density measurements on as-printed Ni-20Cr (10 specimens) and Ni-20Cr-1Y$_2$O$_3$ (28 specimens) as functions of laser power ($P$) and scan speed ($v$) given constant beam radius ($r_B$ = 20 μm), hatch spacing ($\eta$ = 40 μm), and layer thickness ($\beta$ = 20 μm). Broken lines indicate extrapolated isocontours. RAM ODS material required higher laser powers and slower scan speeds to achieve full density.

*3.2. Structural characterization*

To illustrate the effect of alloy composition, **Fig. 3** compares backscatter electron micrographs of Ni-20Cr, ODS Ni-20Cr, and ODS Ni-20Cr alloyed with 1 wt% Al printed using similar parameters. The baseline Ni-20Cr and ODS variant with 0.45 wt% Y$_2$O$_3$ were fully dense and defect-free. The ODS variants with 1 wt% Y$_2$O$_3$ and with 1 wt% Al contained micron-scale oxide slag inclusions (cf. insets, where the bright inclusions were confirmed through EDS measurements to be Y$_2$O$_3$ in **Fig. 3c** and mixed Y-Al-O in **Fig. 3d**). Formation of Y-Al-O slag is consistent with recent reports [6,9] of reactions between Y$_2$O$_3$ and Al during L-PBF, where they form low melting point mixed oxides (YAlO$_3$, Y$_3$Al$_5$O$_{12}$) that rapidly coarsen and agglomerate. Ni-20Cr-1Y$_2$O$_3$ also contained solidification cracks, possibly linked to Y segregation to grain boundaries [6] above a threshold Y$_2$O$_3$ content. All materials featured columnar grains aligned with the build direction except for the variant with 1 wt% Al, which had more equiaxed grains. The ODS variant with 0.34 wt% Al is not shown, but was nearly defect-free, with a columnar grain structure similar to Ni-20Cr-0.45Y$_2$O$_3$. These observations suggest a change in oxide structural evolution behaviors above a threshold Al content.

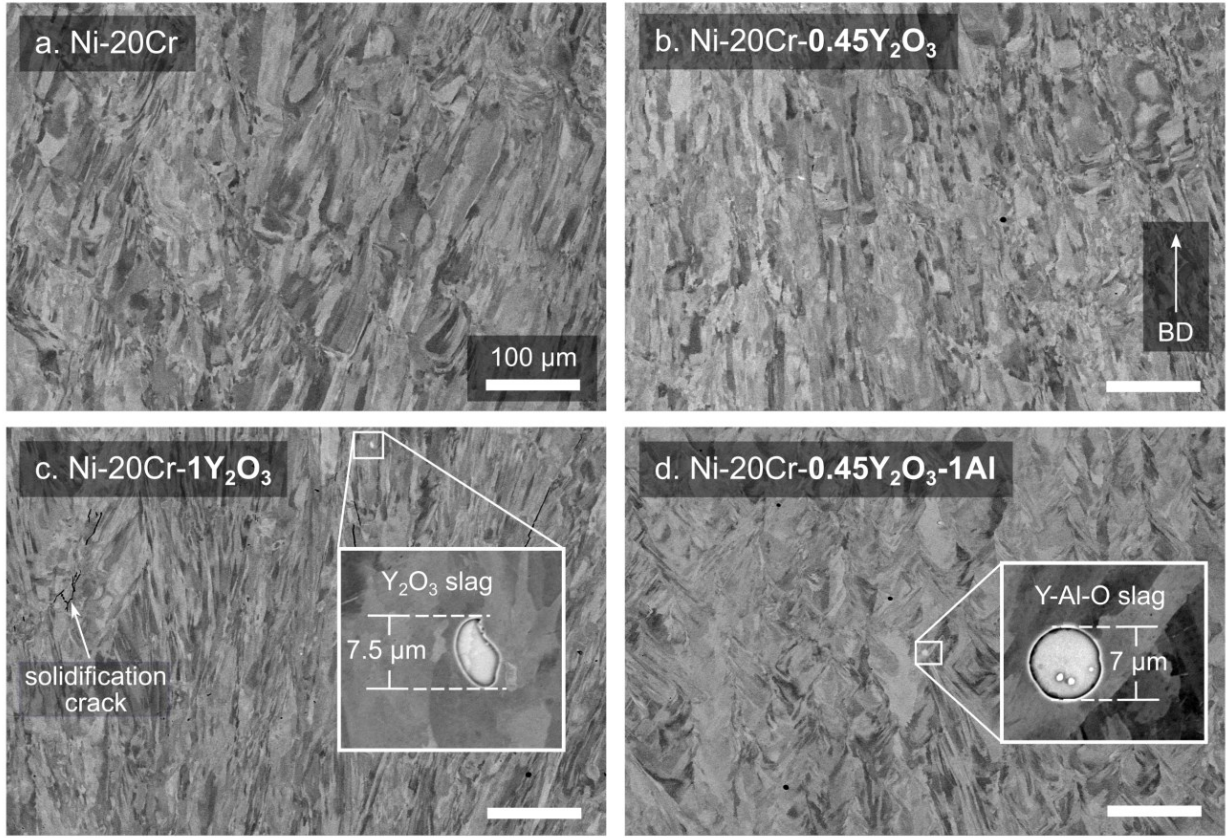

**Fig. 3.** Backscatter electron micrographs of **(a)** Ni-20Cr, **(b)** Ni-20Cr-0.45Y$_2$O$_3$, **(c)** Ni-20Cr-1Y$_2$O$_3$, and **(d)** Ni-20Cr-0.45Y$_2$O$_3$-1Al. The Ni-20Cr and Ni-20Cr-0.45Y$_2$O$_3$ variants are defect-free whereas the high Y$_2$O$_3$ and Al-rich variants contain micron-scale slag inclusions.

The bulk-averaged Y concentration in the 0.45 wt% Y$_2$O$_3$ alloy was 0.12 wt%, as measured with WDS. Assuming Y is sequestered in Y$_2$O$_3$, this corresponds to a retained Y$_2$O$_3$ concentration of 0.16 wt% (0.2 at%), substantially less than the Y$_2$O$_3$ concentration in the powder feedstock. Y$_2$O$_3$ loss was also observed in the other ODS materials. In past work similar Y$_2$O$_3$ loss was attributed to slag formation and material loss [6]. WDS measurements of the alloys doped with Al showed identical Al content in the powder feedstock and as-printed material, indicating a lack of Al volatilization.

**Fig. 4** presents a STEM HAADF micrograph of the as-printed Ni-20Cr-0.45Y$_2$O$_3$ material as well as EDS measurements of an exemplary nanoscale dispersoid. The bright circular features in **Fig. 4** are Y$_2$O$_3$ dispersoids, confirmed through EDS measurements. The dark circular features exhibit no EDS signal and are therefore assumed to be voids from dispersoid pullout during sample preparation. We observed dispersoids with diameters between 8 and 115 nm. Resolving dispersoids smaller than 20 nm was challenging because they had low contrast against the matrix and were obscured by the dense dislocation network.

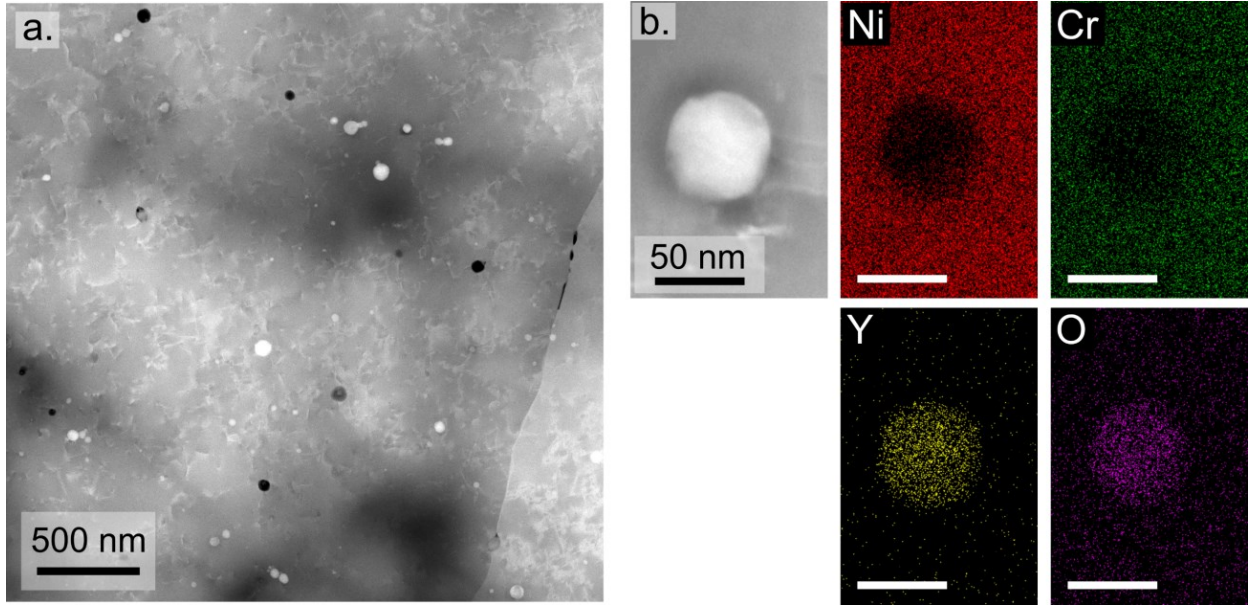

**Fig. 4. (a)** STEM HAADF micrograph of Ni-20Cr-0.45Y$_2$O$_3$. The dense dislocation network obscures dispersoids smaller than ~10 nm. **(b)** STEM EDS measurements of a nanoscale Y$_2$O$_3$ dispersoid.

STEM micrographs of the ODS materials alloyed with Al were qualitatively similar to that in **Fig. 4**, except the dispersoids were slightly larger, suggesting accelerated coarsening kinetics in the presence of Al. EDS measurements of dispersoid chemistry which might indicate reactions between Y$_2$O$_3$ and Al were complicated by the fact that the interaction volume includes the metal matrix surrounding the particle. Thus, to estimate the atomic ratio of Al to Y in each dispersoid, we used the following expression to subtract the matrix signal:

$$\frac{[Al]}{[Y]} = \frac{[Al]_d}{[Y]_d} - \frac{[Al]_m}{[Y]_d}\frac{[Ni]_d}{[Ni]_m}, \qquad (1)$$

where terms in brackets are atomic concentrations measured through EDS, and the subscripts $d$ and $m$ refer to measurements of the embedded dispersoid and of the baseline metal matrix at a dispersoid-free location. The atomic ratio of Al to Y measured this way is plotted against dispersoid size in **Fig. 5**. The dispersoids in the 0.3 wt% Al alloy have negligible Al content. By contrast in the 1 wt% Al alloy there is significant Al enrichment, with the Al content increasing with dispersoid size. The largest dispersoids have an Al to Y ratio of ~0.2. For comparison, the Al to Y ratio is 0.33 at the Y$_2$O$_3$/Y$_4$Al$_2$O$_9$ eutectic. These findings indicate that Al can react with nanoscale dispersoids, potentially having an important role in the earliest stages of oxide structural evolution.

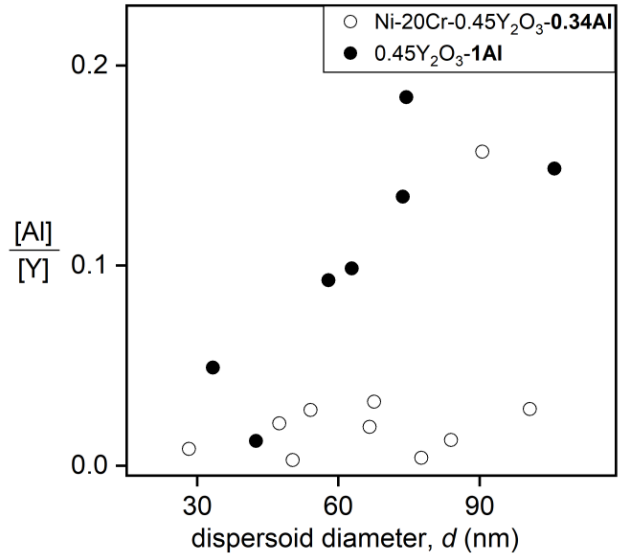

**Fig. 5.** Dispersoid composition vs. diameter for ODS alloys with varying Al content. The dispersoids were only slightly Al enriched in the 0.3 wt% Al alloy, whereas the Al content increased with dispersoid size in the 1 wt% Al alloy.

The above results reveal three key features of oxide structural evolution during L-PBF of ODS alloys, informing the models in **Sec. 4**. First, the as-printed material contains spherical nanoscale dispersoids that are uniformly dispersed throughout the metal matrix. Second, the volume fraction of these dispersoids is less than anticipated based on the feedstock composition. This is potentially because of slag formation and spatter. Third, Al reacts with $Y_2O_3$, concentrating in larger $Y_2O_3$ dispersoids and promoting slag formation. The latter finding further emphasizes the important role of material chemistry in oxide structural evolution processes.

*3.3. SANS measurements of dispersoid size distribution*

Two major limitations of using electron microscopy to measure the dispersoid size distribution are: (i) the small sample volumes and (ii) the difficulty of resolving sub-10 nm dispersoids. Here we overcome these challenges using SANS, which benefits from large interaction volumes (~5 mm$^3$) and can resolve dispersoids over the complete anticipated size range. **Fig. 6** shows the SANS spectrum collected from a Ni-20Cr-0.45$Y_2O_3$ specimen, printed using $P = 140$ W, $v = 1.2$ m/s. The form of the spectrum within the $Q$-range 0.02-0.6 nm$^{-1}$ indicates a nanoscale second phase, with diameters between 7 and 100 nm, in line with our STEM measurements.

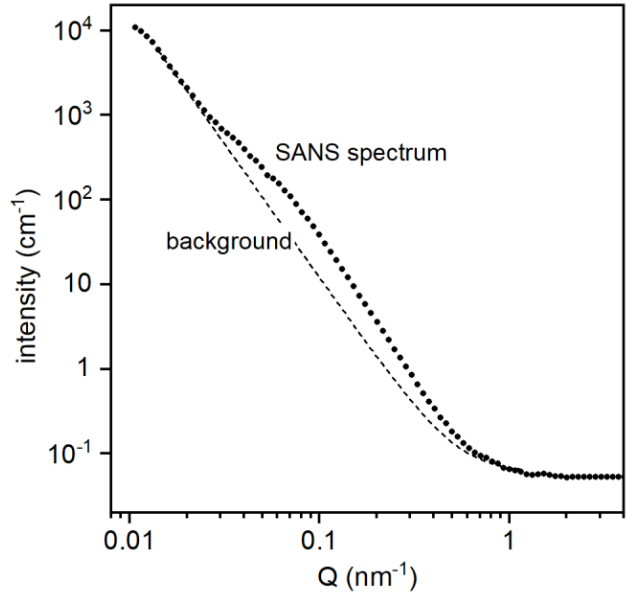

**Fig. 6.** SANS spectrum from L-PBF Ni-20Cr-0.45Y$_2$O$_3$ ($P$ = 140 W, $v$ = 1.2 m/s).

**Fig. 7** shows the dispersoid size distribution calculated using the SANS spectrum in **Fig. 6**. For comparison purposes we have also included STEM measurements of the dispersoid size distribution and literature data for a similar alloy (Ni-20Cr-1Y$_2$O$_3$) processed under comparable conditions ($P$ = 200 W, $v$ = 0.9 m/s, $r_B$ = 50 µm) [6]. The SANS distribution is approximately log-normal, with a mean dispersoid size of 21 nm. The volume percent of Y$_2$O$_3$ dispersoids is 0.24%, in line with the WDS measurement of 0.27 vol%, confirming Y is present as Y$_2$O$_3$. While there is good agreement between our STEM measurements and the data in [6], comparing the SANS and STEM dispersoid size distributions shows that STEM undercounts dispersoid sizes smaller than 25 nm. This systematic error in the STEM data arises from the aforementioned challenges with resolving dispersoids (cf. **Sec. 3.2**). As a result, the STEM measurements yield a larger mean dispersoid size (21 nm in SANS vs. 28 nm in STEM) and lower dispersoid number density (230 vs. 120 µm$^{-3}$). Additionally, SANS has a statistically significant sampling volume of order 5 mm$^3$, which corresponds to roughly 10$^{11}$ dispersoids, roughly nine orders of magnitude greater than the sample sizes in STEM.

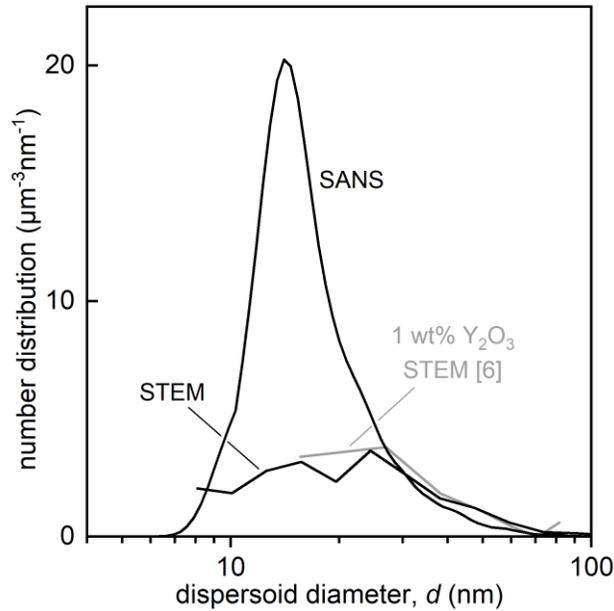

**Fig. 7.** Dispersoid size distribution in Ni-20Cr-0.45Y$_2$O$_3$ ($P$ = 140 W, $v$ = 1.2 m/s). The SANS size distribution is compared to STEM stereological measurements collected here and in [6] on a similar alloy.

**Fig. 8** compares the SANS dispersoid size distribution of specimens with varying Al and Y$_2$O$_3$ content, all printed with the same parameters ($P$ = 140 W, $v$ = 1.2 m/s). Increasing the Al content to 1 wt% caused the mean dispersoid size to increase from 21 to 27 nm and the volume percent of dispersoids to decrease from 0.24 to 0.15 vol%. Al additions decreased dispersoid concentration across the full range of dispersoid sizes, and in the 1 wt% Al material, essentially eliminated sub-20 nm dispersoids. These missing dispersoids were likely incorporated into slag. Increasing the Y$_2$O$_3$ concentration also resulted in a larger mean dispersoid size. The mean dispersoid size of the 1 wt% Y$_2$O$_3$ material was 33 nm, nearly twice that of the baseline material with 0.45 wt% Y$_2$O$_3$. Interestingly, both materials had the same final Y$_2$O$_3$ volume percent of 0.24%.

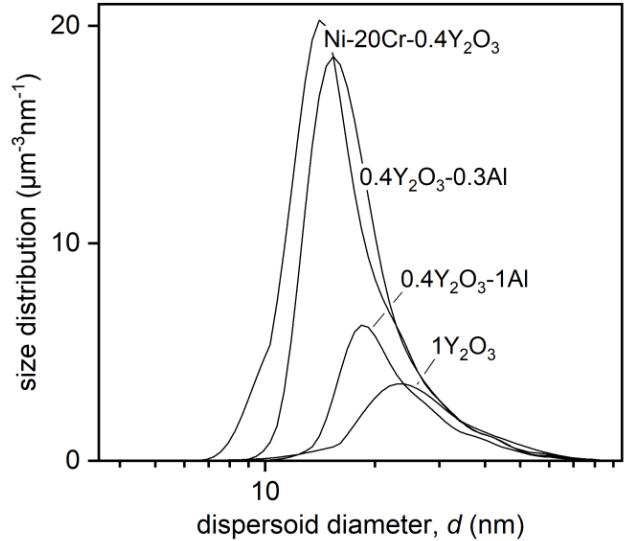

**Fig. 8.** Dispersoid size distributions in alloys with varying $Y_2O_3$ and Al content ($P$ = 140 W, $v$ = 1.2 m/s), showing an increase in mean dispersoid size with alloying additions.

**Fig. 9** shows the effects of beam power and scan speed on the dispersoid size distribution. The trends in **Fig. 9** are broadly consistent with those reported in the open literature: increasing the beam power and decreasing the scan speed increased the dispersoid size and lowered the dispersoid number density. $Y_2O_3$ content varied from 0.17 to 0.25 vol% and was generally lower for parameter combinations that prolonged the melt time.

For comparison **Fig. 9a** includes SANS measurements of the dispersoid size distribution in wrought MA754 (Ni-20Cr-0.34Al-0.43Ti-0.6$Y_2O_3$), an ODS alloy with a similar composition to the alloys of present interest. MA754 had a mean dispersoid size of 19 nm, dispersoid number density of 2000 $\mu m^{-3}$, and dispersoid volume fraction of 1.45%. The SANS measurement of the MA754 dispersoid size distribution is in good agreement with past small angle X-ray scattering measurements [9], thus validating the SANS measurements. The L-PBF ODS material printed with the fastest beam speed (1.6 m/s) had a mean dispersoid size of 16 nm, a dispersoid number density of 600 $\mu m^{-3}$, and a dispersoid volume fraction of 0.25%. Thus, while the L-PBF and wrought materials can achieve similar dispersoid sizes, the number density and volume fraction of dispersoids in the L-PBF material are an order of magnitude lower, pointing towards a smaller strengthening increment in L-PBF materials.

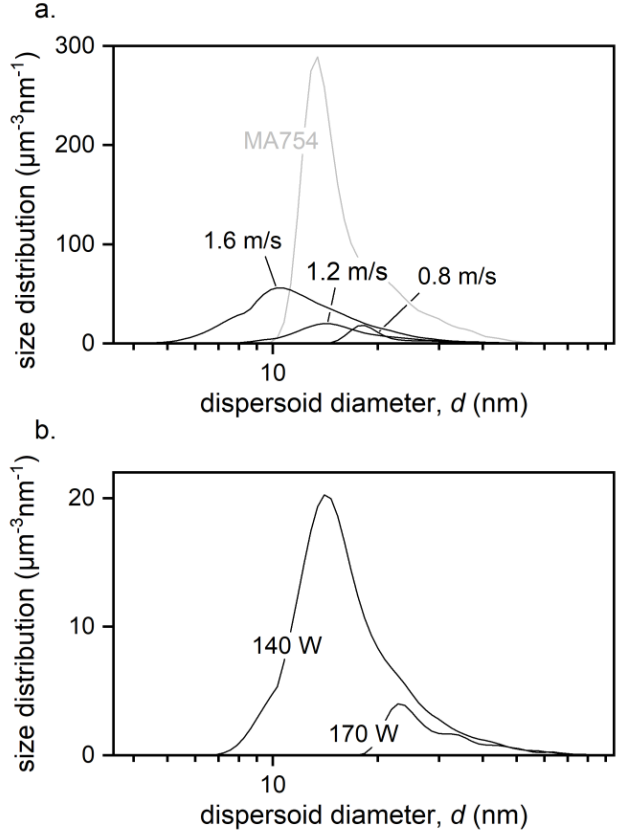

**Fig. 9.** Dispersoid size distributions in Ni-20Cr-0.45Y$_2$O$_3$ printed **(a)** with varying scan speed but fixed power (140 W), and **(b)** with varying power but fixed scan speed (1.2 m/s).

*3.4. Overview of experimental observations and comparison with literature data*

We can understand our experimental results in the context of recent work on L-PBF ODS Ni-base alloys (cf. [6,7,10,12,13,29]) using **Fig. 10**, an L-PBF process diagram which aggregates our data and that from the open literature. Generally, different studies use different processing conditions, L-PBF printers, etc. To facilitate direct comparisons, **Fig. 10.** has axes of dimensionless laser power $\tilde{P}$ and scan speed $\tilde{v}$, defined as [30]:

$$\tilde{P} = \frac{AP}{r_B \lambda (T_m - T_0)} \text{ and} \quad (2)$$

$$\tilde{v} = \frac{v \rho c_p r_B}{\lambda}, \quad (3)$$

where $r_B$ is the beam radius, $A$ is absorptivity (assumed to be 0.3 [31]), $\lambda$ is thermal conductivity, and $\rho c_p$ is volumetric heat capacity (3 × 10$^6$ J/m$^3$K). The aggregated data is clustered in $\tilde{P}$-$\tilde{v}$ space, near the transition between melting and vaporization [30], where keyhole melting dominates [32].

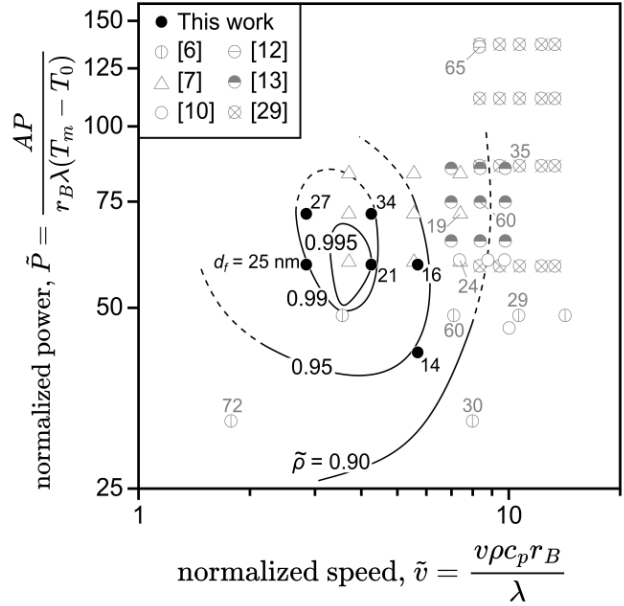

**Fig. 10.** L-PBF process diagram for Ni-base ODS alloys, with axes of dimensionless beam power ($\tilde{P}$) and scan speed ($\tilde{v}$). Labels indicate mean dispersoid size. Isocontours of relative density from **Fig. 2** are overlaid.

The labels in **Fig. 10** indicate mean dispersoid size. Our measurements of dispersoid size are the smallest reported to date, likely because past studies assessed dispersoid size using electron microscopy [6,7,10,12,13,29], while we used SANS, which has the advantages mentioned above. The lack of clear trends in the aggregated dispersoid size data reflects variations in hatch spacing and alloy chemistry. However, in the two datasets that contain multiple dispersoid size measurements – [6] and our study – it is clear that increasing scan speed and decreasing beam power result in smaller dispersoids.

Ultimately, scan speed and beam power must also fully densify the feedstock, thus limiting the practical processing window and the minimum achievable dispersoid size. To illustrate this point, we have overlaid **Fig. 10** with the relative density isocontours of **Fig. 2**. Comparing these datasets shows that ~20 nm is the smallest achievable dispersoid size in fully-dense as-printed material. However, there may be multi-step processing routes that unlock smaller dispersoids; e.g., printing 95% dense material with a 14 nm dispersoid size, then eliminating residual porosity through post-process hot isostatic pressing.

Our experimental results provide direct quantitative evidence of several trends previously speculated in the open literature. They also highlight key differences between wrought ODS alloys and their L-PBF counterparts that may limit the creep resistance of L-PBF ODS alloys. Specifically, wrought and L-PBF materials have similar dispersoid sizes, but the dispersoid density of wrought materials is an order of magnitude higher. Furthermore, wrought ODS materials such as MA6000 can benefit from γ'-strengthening through Al alloying additions, whereas L-PBF ODS alloys develop slag when alloyed with Al, even in small quantities around 1 wt%. In the following section we develop a quantitative physics-based model of dispersoid structural evolution to identify strategies for achieving wrought-like dispersoid structures via melt-based additive manufacturing.

## 4. Modeling dispersoid structural evolution

### 4.1. *Scaling analysis*

Developing a quantitative model of oxide structural evolution requires knowledge of the dominant oxide coarsening mechanisms, which vary with processing conditions, alloy chemistry, and dispersoid size. To determine which mechanisms dominate under specific conditions, we turn to scaling analysis of momentum and mass transport. We first assess flow behaviors within the melt pool and in the vicinity of the dispersoids by computing their respective Reynolds numbers, $Re = \frac{\rho v l}{\mu}$, where $\mu$ is dynamic viscosity, $\rho$ is fluid density, and $l$ and $v$ are characteristic lengthscale and fluid velocity. The characteristic lengthscale and velocity of the melt pool are the melt pool dimensions, defined in **Fig. 13,** and the Marangoni-driven fluid velocity $v_{melt}$. The present calculations assume a melt pool length ($l$) of 240 μm and depth ($h$) of 40 μm, typical values in L-PBF. The characteristic fluid velocity of surface tension-driven flow is (cf. **Appendix A**):

$$v_{melt} = \frac{\Delta T_{VM}(d\gamma_{LG}/dT)}{4\mu}\frac{h}{l}, \tag{4}$$

where $\Delta T_{VM}$ is the difference between the alloy vaporization and melting temperatures, $d\gamma_{LG}/dT$ is the surface tension coefficient, and $h/l$ is the melt pool aspect ratio. Inserting values for L-PBF of Ni-20Cr from **Table 2** into **Eq. 4** gives $v_{melt}$ = 5 m/s. This value gives a $Re$ number of order ~2000, suggesting that at least in the tail of the melt pool the flow is borderline laminar.

Dispersoids may move relative to the surrounding flow due to the large density difference between the metal matrix (~8-9 g/cm$^3$ for Ni-base superalloys) and the dispersoid (5 g/cm$^3$ for Y$_2$O$_3$). Relative motion of the dispersoids is associated with a local disturbance in the flow field which transitions from the dispersoid velocity, $v_p$, at the interface (i.e., no slip) to $v_{melt}$ in the far field. The characteristic lengthscale and velocity of the local $Re$ number near an oxide are the oxide diameter, $d$, and a local relative velocity between the oxide and the bulk fluid velocity, $v_{rel} = v_p - v_{melt}$, which depends on the flow regime (e.g., creeping vs. inviscid). If we assume that the nanoscale dispersoids experience creeping flow, their $v_{rel}$ can be estimated using Stokes law (cf. [33], **Appendix A**):

$$v_{rel} = \frac{\Delta\rho v_{melt} d^2}{9\mu t_{res}}, \tag{5}$$

where $\Delta\rho$ is the density difference between the oxide and the melt, and $t_{res}$ is the residence time of a fluid element (~100 μs for L-PBF). Values of local $Re$, calculated using **Eq. 5,** are plotted as a function of oxide diameter in **Fig. 11**, which shows that $Re$ is less than unity for oxide particles smaller than 8 μm and is of order $10^{-9}$–$10^{-6}$ for nanoscale dispersoids, justifying our earlier assumption of creeping flow.

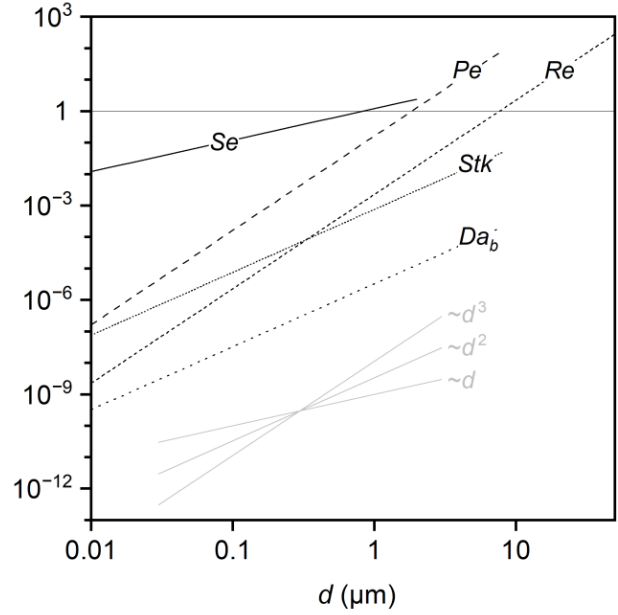

**Fig. 11.** Dimensionless numbers that characterize momentum and mass transport as a function of dispersoid size $d$. These parameters indicate that nanoscale oxide dispersoids smaller than 1 μm experience creeping flow and grow through diffusion-mediated mass transport mechanisms.

Recognizing that nanoscale dispersoids experience creeping flow, we assess the significance of coarsening via mechanical impingement by computing two additional dimensionless quantities: the buoyancy number ($Da_b = \frac{4t_{res}\Delta\rho g d^2}{27\mu h}$) which compares residence time to the particle floatation timescale, and the Stokes number ($Stk = \frac{\Delta\rho v_{melt} d^2}{18\mu l}$) which compares the timescale for a suspended particle to match the velocity of the surrounding flow to the melt pool flow timescale ($\frac{v_{melt}}{l}$). **Fig. 11** shows that the buoyancy and Stokes numbers are both much less than unity for dispersoids smaller than 5 μm, indicating that small particles will follow streamlines and avoid agglomeration via floatation. In addition, the time reversibility of Stokes flow guarantees particles that are initially separated will remain separated. Taken together, these results indicate that oxide particle coarsening via impingement of particles is unlikely for oxides smaller than 5 μm.

We also compute dimensionless quantities which provide insight into diffusion-mediated structural evolution mechanisms. To assess the significance of mass-transport via advection, we calculate the Peclet number ($Pe = \frac{v_{rel} d}{2D}$), where $D$ is the chemical diffusivity of dissolved oxide (of order $8 \times 10^{-9}$ m$^2$/s [34]). **Fig. 11** shows $Pe \ll 1$ for dispersoids smaller than 800 nm, indicating advection is insignificant for nanoscale dispersoids. Further, the separation number ($Se = \frac{1}{4\sqrt{Dt_{res}}} \frac{d}{V_{f,ox}^{1/3}}$), i.e., the ratio of the mean dispersoid separation distance to the diffusion lengthscale, is less than unity for oxides smaller than 100 nm, establishing that nanoscale dispersoids are sufficiently close for local diffusive mass transport between neighboring particles.

Considering the various dimensionless numbers in **Fig. 11**, we conclude that nanoscale dispersoids experience creeping flow, with limited motion relative to the local environment, and evolve through diffusion-mediated structural evolution processes. By contrast, micron-scale oxides experience additional coarsening mechanisms. Oxides larger than 2 μm have Peclet numbers greater than unity and grow primarily through advection. Advection is generally faster than diffusion and may accelerate coarsening in oxides that have grown to this threshold size [33]. Oxides larger than 10 μm are no longer bound by creeping flow and may coarsen via impingement or by agglomerating at the surface, further accelerating coarsening. Our model focuses on the 10-100 nm dispersoids (**Sec. 4**), where diffusion is the sole mass transfer mechanism and which impart the most creep resistance. Based on the dispersoid evolution behavior in this size range, however, we predict the set of conditions that permit dispersoids to grow to the accelerated coarsening regime and form slag (**Sec. 5**).

**Table 2.** Thermophysical inputs to scaling analysis and dissolution zone model. Ranges are given for temperature-dependent properties.

| Parameter | | value | | reference |
|---|---|---|---|---|
| Ni-20Cr | | | | |
| $T_m$ | melting temperature (K) | 1673 | | [35] |
| $T_b$ | boiling temperature (K) | 3003 | (Ni) | – |
| $\rho$ | density (g/cm$^3$) | 8.1-8.5 | (solid) | [36] |
| | | 7.5-7.7 | (molten) | [36] |
| $c_p$ | specific heat capacity (J/kgK) | 440-737 | (solid) | [37,38] |
| | | 840 | (molten) | [38] |
| $\lambda$ | thermal conductivity (W/mK) | 14.3-38.0 | (solid) | [37,39] |
| | | 31.7-43.7 | (molten) | [38] |
| $\mu$ | dynamic viscosity (mPa s) | 4.1-5.7 | | [40] |
| $\gamma_{LG}$ | surface tension (N/m) | 1.67-1.81 | | [41] |
| $h_f$ | latent heat of fusion (kJ/kg) | 280 | | [38] |
| $V_m$ | molar volume (m$^3$/mol) | $7.5 \times 10^{-6}$ | (molten) | – |
| Y$_2$O$_3$ | | | | |
| $\gamma_{SL}$ | interfacial energy (N/m) | 2.0 | (with Ni-20Cr) | [23,42] |
| $D_0$ | diffusion coefficient (m$^2$/s) | $5.6 \times 10^{-7}$ | | [34] |
| $E_D$ | diffusion activation energy (J/mol) | $8.2 \times 10^4$ | | [34] |
| $\rho_{ox}$ | density (g/cm$^3$) | 5.0 | | – |
| $V_{m, ox}$ | molar volume (m$^3$/mol) | $9.0 \times 10^{-6}$ | | – |

### 4.2. *Model framework*

Motivated by the preceding scaling analysis we develop a modeling framework, depicted in **Fig. 12**, for predicting the oxide structure in L-PBF ODS alloys. First, we use a reduced order model, calibrated with computational fluid dynamics, to compute the time-temperature trajectories of fluid elements throughout the melt pool. Next these time-temperature trajectories are fed into numerical simulations which track dispersoid size and number density as they evolve through dissolution, nucleation, growth, and coarsening over a single melt cycle. Two other key inputs for the numerical simulations are temperature-dependent solubility and the free energy change of the oxide upon precipitation, which come from a thermochemical model which takes into account alloy composition. For processing conditions that fully dissolve the dispersoids in the feedstock, the dispersoid size and number density after the single melt cycle are representative of the bulk as-printed dispersoid structure and are thus the desired output. Processing conditions that only dissolve a fraction of the dispersoids in the feedstock are assessed for likelihood of slag retention through a multi-track model, i.e., the dissolution zone model, developed in **Sec. 5**.

The present framework implements simplified, computationally efficient models that capture the essential physics. For instance, we use classical mean field theory to describe the particle growth kinetics, rather than population tracking models like KWN [43]. This approach sacrifices some precision for the ability to capture major trends and allows for the high-throughput parametric calculations needed for the dissolution zone model in **Sec. 5**.

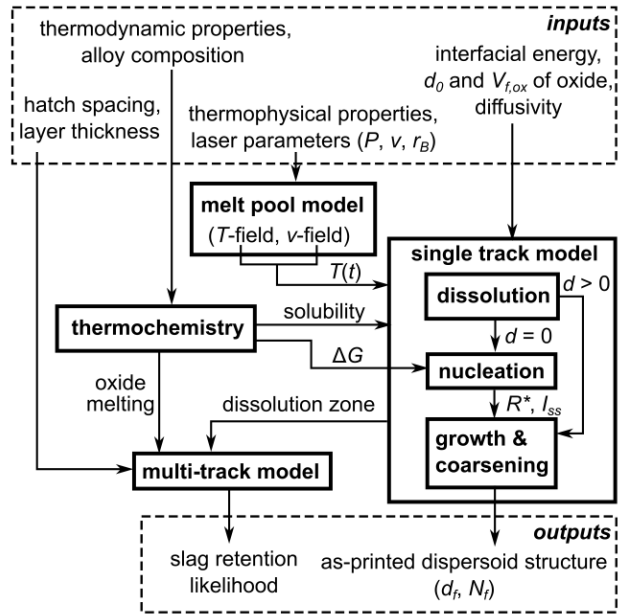

**Fig. 12.** Modeling framework.

4.3. *Thermal excursions in the L-PBF melt pool*

We use computational fluid dynamics (CFD) to obtain the thermal excursion of fluid elements as they translate with the flow, replicating the time-temperature history that oxides experience from initial melting through solidification. Detailed modeling of the thermal excursion is critical because quantities that govern dispersoid evolution – e.g., diffusivity, solubility, nucleation rate – vary strongly with temperature. After an initial analysis following from the CFD model, simplifying assumptions are used in **Sec. 4.7** to develop an analytical reduced order model of the temperature field for high-throughput calculations.

Single-track L-PBF simulations were performed using the commercial CFD package Flow-3D. Temperature-dependent thermophysical properties of Ni-20Cr used for the simulations are listed in **Table 2**. The low oxide concentration in the present ODS alloys (~0.24 vol%) is expected to have a minor impact on the melt pool flow behavior. Most thermophysical properties of the melt are insensitive to nanoscale dispersoids [44,45], except for viscosity. For example, Qu et al. observed a 4-15X increase in the viscosity of Al6061 with the addition of 1.8-4.4 vol% TiC nanoparticles, leading to lower melt-pool fluid velocities [46]. However, the dispersoid volume fraction in the present study is an order of magnitude lower, resulting in only a ~20% increase in viscosity [45]. Moreover, oxides influence bulk viscosity only when present as a dispersed second phase, which exists only in cooler regions of the melt pool (see **Sec. 4.4**) where convection is less significant. Thus, omitting dispersoid effects on viscosity maintains the underlying physics and is unlikely to significantly affect the results.

Equations for heat conduction, mass conservation, and momentum conservation were solved numerically using implicit first order schemes [47,48]. The free surface at the top of the melt pool is tracked using the split Lagrangian method [49]. Additional key physics modeled include: energy loss and recoil pressure from vaporization [50]; latent heat effects during melting and solidification; and Marangoni convection, using temperature-dependent surface tension data of Ni-20Cr [41].

The simulation domain, which encompassed half of the symmetrical melt track (see **Fig. 13**), was 1.2 mm long, 120 μm wide, and 120 μm deep, with a 20 μm gap above the fluid region to allow surface deformation. The center plane of the melt track was assigned a symmetry boundary condition. The top boundary was set to maintain constant ambient pressure and temperature. All remaining surfaces were assigned wall boundary conditions with constant ambient temperature. The mesh contained 205,000 cells and was graded with fine cells (1 μm) at the center of the melt track, where there are large fluid velocities and thermal gradients, transitioning to coarse cells (10 μm) at the periphery. The simulation was run for 850 μs. After approximately 400 μs, the melt pool reaches steady state. The steady-state melt boundary, temperature field, and velocity field were used to model dispersoid evolution.

**Fig. 13** presents time series images of a melt pool during L-PBF with laser parameters $P$ = 140 W, $v$ = 1.2 m/s, and $r_B$ = 20 μm. The temperature field is non-uniform, exceeding the Ni-20Cr boiling temperature directly under the laser. The melt pool is defined as the region where the temperature was higher than the Ni-20Cr solidus ($T_m$ = 1673 K). For this and subsequent analyses, we assume a single melting temperature, equal to the solidus because of the narrow melting range of Ni-20Cr (1673–1690 K [35]). The steady-state melt pool is 244 μm long, 54 μm wide, and 40 μm deep. The Eagar-Tsai model for conduction-mode melting by a Gaussian heat source [51,52] predicts a shorter, wider melt pool (175 μm long, 82 μm wide, 35 μm deep). This discrepancy arises because the CFD model accounts for Marangoni-driven flow while Eagar-Tsai does not.

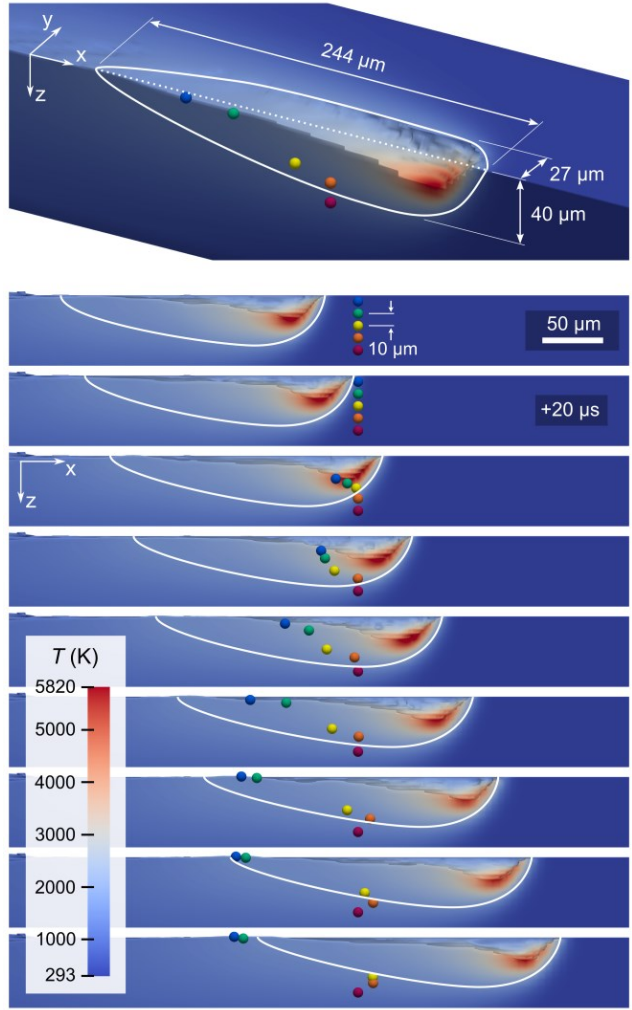

**Fig. 13.** Time-series images from a CFD simulation of the melt pool during L-PBF of Ni-20Cr with laser parameters $P = 140$ W, $v = 1.2$ m/s, and $r_B = 20$ μm. The colored spheres are tracer particles for extracting the time-temperature profile of fluid elements within the melt pool.

A thermal gradient on the melt pool surface, of order $10^7$ K/m, creates a gradient in surface tension that accelerates fluid away from the laser spot, towards the melt pool periphery. The resulting velocity field is shown in **Appendix A** (see **Fig. A.1b-c**), which shows that the dominant fluid motion is along the x-axis, opposite the scan direction, with a maximum velocity of 4.1 m/s and mean velocity ($\bar{v}_x$) of 0.45 m/s. The maximum velocity agrees well with our analytical order-of-magnitude estimate of 5 m/s (cf. **Appendix A**). There is also motion in the transverse directions, with y- and z- components attaining maximum speeds of 4.1 m/s and 8.1 m/s, respectively. However, transverse motion is only significant near the vapor depression. As a result, the average velocities along the y- and z-directions are relatively slow – 0.04 and 0.1 m/s, respectively.

**Fig. 13** shows tracer particles placed at various depths along the melt pool symmetry plane. These tracer particles were used to measure thermal excursions of fluid elements traversing the melt pool. The particles primarily move in the x-direction, with the largest displacements near the surface. The tracers exited the melt pool via solidification before re-entering the convection cell, experiencing a single thermal excursion during the melt cycle. In general, full recirculation at the melt-pool scale is unlikely in L-PBF because the average residence time ($l/v_{\text{laser}} \approx 200$ μs) is less than the time required to complete a single convection cycle ($2\,l/\bar{v}_x \approx 1$ ms). A simulation performed with a large array of tracer particles (cf. Supplementary Materials) shows that while circulatory mixing occurs in the y-z plane, individual particles do not recirculate. This is consistent with the experimental results in [53]: although the maximum fluid velocity is of the same order as the laser scan velocity near the vapor depression, the median velocity is significantly slower, thus precluding full recirculation of individual fluid elements. Consequently, typical L-PBF thermal excursions are expected to feature a single peak in temperature, in contrast with the oscillating thermal cycles observed in large, slow-moving weld pools [22].

The thermal excursions experienced by the tracer particles are shown in **Fig. 14a**. The labels indicate the initial depth of each tracer particle normalized by the beam radius, $\tilde{z} = \frac{z}{r_B}$. While the thermal excursions are qualitatively similar to those expected from the Eagar-Tsai solution (see comparison in **Appendix C, Fig. C.1a**), there are two key differences. First, for trajectories near the surface, the peak of the thermal excursion in the CFD simulations is narrower due to rapid acceleration from the laser spot by the surface tension-driven flow. Second, trajectories underneath the surface experience higher temperatures in CFD because of heat redistribution via convection as well as depression of the melt pool from the vapor recoil pressure.

The wide temperature range in the melt pool reinforces the need for a detailed treatment of dispersoid evolution at the sub-melt pool scale. At the same time, the present observations also motivate a reduced order heat conduction model modified to reproduce the time-temperature profiles from CFD. This is developed in **Sec. 4.7**.

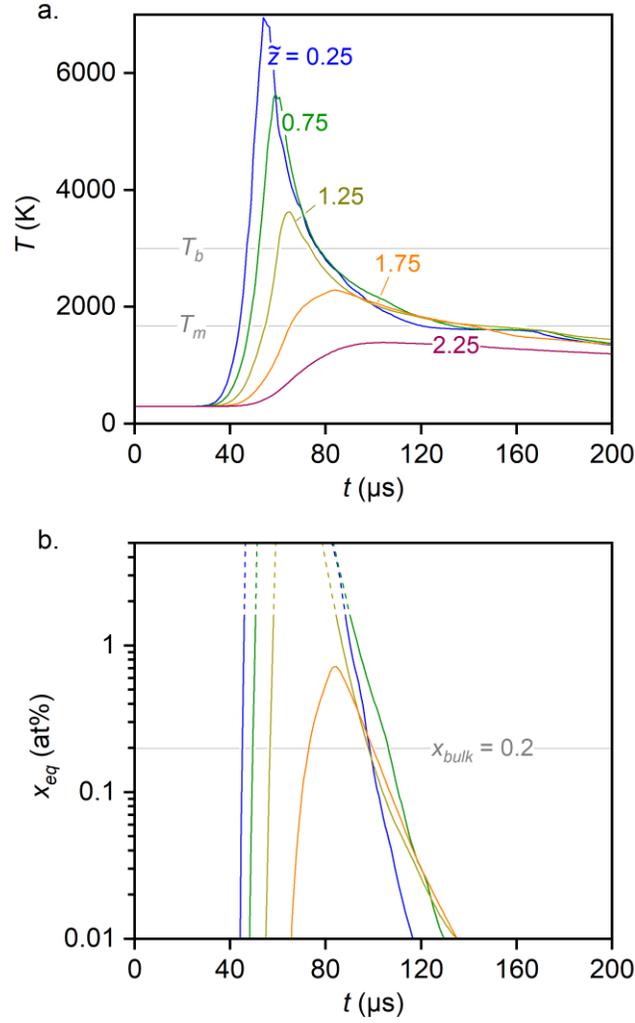

**Fig. 14. (a)** Thermal excursions and **(b)** resulting $Y_2O_3$ solubility for tracer particles shown in **Fig. 13** ($P = 140$ W, $v = 1.2$ m/s).

### 4.4. Oxide solubility

Another key input for modeling dispersoid evolution is the solubility of the oxide in the alloy melt. Assuming thermodynamic equilibrium at the oxide-metal interface, the solubility of $Y_2O_3$ follows from the reaction

$$\langle Y_2O_3 \rangle \leftrightarrow 2[Y]_{Ni} + 3[O]_{Ni}, \tag{6}$$

where angled brackets indicate pure states and square brackets indicate chemical species dissolved in the alloy matrix. The free energy change associated with this reaction is

$$\Delta G^\circ(T) = -\Delta G^{\circ f}_{Y_2O_3} + 2\,\Delta G^{\circ sol}_Y + 3\,\Delta G^{\circ sol}_O, \tag{7}$$

where the temperature-dependent energies of the constituent reactions are available in literature: $\Delta G^{\circ f}_{Y_2O_3}$ is the energy of formation of $Y_2O_3$ [17,54], $\Delta G^{\circ sol}_Y$ is the heat of solution of Y in Ni [55], and $\Delta G^{\circ sol}_O$ is the heat of solution of O in Ni [56]. The equilibrium constant for this reaction is

$$\frac{a_Y^2 a_O^3}{a_{Y_2O_3}} = K_c(T) = \exp\left(-\frac{\Delta G^\circ}{RT}\right), \tag{8}$$

where $a$ is the activity. The activity of pure $Y_2O_3$, $a_{Y_2O_3}$, is approximated as unity. The activities of dissolved Y and O ($a_Y$, $a_O$) are approximated as their respective mole fractions $x_Y$ and $x_O$ (i.e., activity coefficients $f_Y$ and $f_O$ are set to unity) due to the lack of thermodynamic data for the Ni-Y-O ternary over the temperature range of interest.

Following these approximations, **Eq. 8** simplifies to $x_Y^2 x_O^3 = K_c(T)$. Stoichiometric dissolution of Y and O dictates that $x_Y = \frac{2}{3} x_O$, providing the constraint necessary to solve **Eq. 8**:

$$x_Y = \left(\frac{8}{27} K_c(T)\right)^{1/5} \text{ and } x_O = \left(\frac{9}{4} K_c(T)\right)^{1/5} \tag{9}$$

The quantity $x_{eq} = x_Y + x_O = 5\left(\frac{K_c(T)}{108}\right)^{1/5}$ is the solubility of $Y_2O_3$, i.e., the combined mole fraction of Y and O atoms dissolved in the alloy matrix at equilibrium. For the discussion of oxide dissolution, growth, and coarsening kinetics that follows, we treat $Y_2O_3$ as a monatomic substance, with Y and O having the same diffusivity and proportional solubility. This simplifying assumption is warranted given uncertainty in other model inputs (e.g., surface energy, chemical diffusivity) at the melt pool temperatures.

$Y_2O_3$ solubility is plotted against temperature in **Fig. 15**, which shows oxide solubility is negligible (< 0.01 at%) at the Ni-20Cr melting temperature then rises sharply to the bulk $Y_2O_3$ content of 0.2 at% at 2080K. At 2345 K, the solubility of $Y_2O_3$ exceeds 1 at%, the maximum $Y_2O_3$ content in conventional ODS alloys. When the peak melt pool temperature exceeds this threshold, dissolution of oxide dispersoids is thermodynamically feasible. In the case of L-PBF ODS Ni-20Cr, 2345 K is well below the peak temperatures predicted by our CFD simulations, suggesting dispersoids can fully dissolve within the melt.

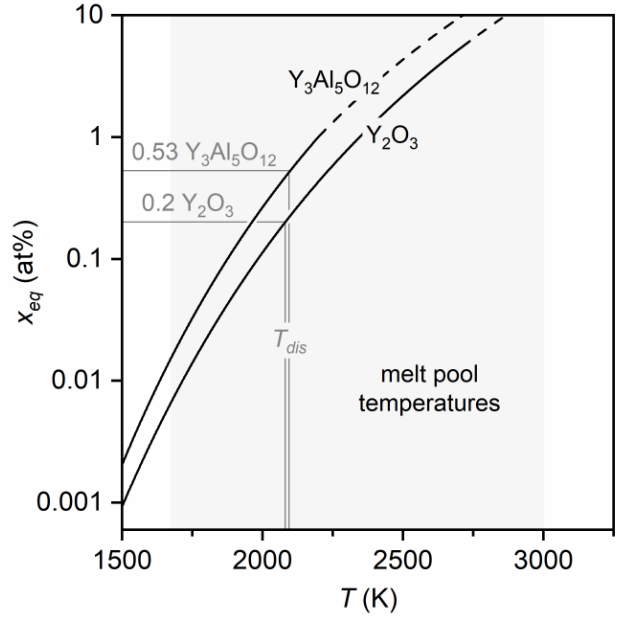

**Fig. 15.** Temperature-dependent concentrations of $Y_2O_3$ and $Y_3Al_5O_{12}$ in liquid Ni. Broken lines indicate where the oxide is molten. The solubility curve of $YAlO_3$ is essentially the same as that of $Y_3Al_5O_{12}$. The shaded region indicates the temperature window bounded by the melting and boiling temperatures of Ni-20Cr. The dissolution temperatures ($T_{dis}$) of 0.2 at% $Y_2O_3$ and 0.53 at% $Y_3Al_5O_{12}$ are 2080 K and 2095 K, as indicated by the vertical lines.

The temperature for full dissolution ($T_{dis}$) is several 100K below the $Y_2O_3$ melting temperature (2711 K) [17]. Nanoscale oxides, which dissolve quickly, should therefore dissolve while still solid. However, larger oxides, whose dissolution rate is diffusion-limited, may melt prior to dissolving, as shown analytically in **Appendix B**. This is consistent with the solidification structures in slag and other evidence of $Y_2O_3$ melting reported in our earlier work [10]. **Sec. 4.5** discusses the kinetics of dissolution and whether an oxide of a given size fully dissolves.

To assess the effect of Al, the same solubility calculations were performed, supposing that the $Y_2O_3$ in the alloy reacted completely with Al and O to form either $YAlO_3$ or $Y_3Al_5O_{12}$. The equilibria for oxide dissolution become

$$\langle YAlO_3 \rangle \leftrightarrow [Y]_{Ni} + [Al]_{Ni} + 3[O]_{Ni} \text{ and} \tag{10}$$

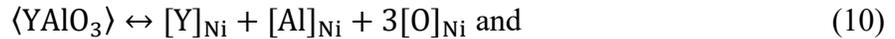

$$\langle Y_3Al_5O_{12} \rangle \leftrightarrow 3[Y]_{Ni} + 5[Al]_{Ni} + 12[O]_{Ni}. \tag{11}$$

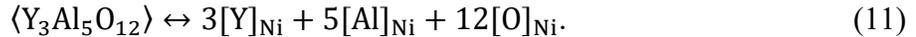

These mixed Y-Al-O phases are less stable than $Y_2O_3$, with lower melting temperatures (cf. **Fig. 15**) and free energies of dissolution. The analysis given by **Eqs. 7-9** is repeated with the added Al component. The resulting solubilities are compared with that of $Y_2O_3$ in **Fig. 15**, which shows that alloying with Al doubles the solubility of the Y-Al-O phases as compared to $Y_2O_3$. This increased solubility accelerates several diffusion-controlled processes, with important implications for dispersoid structure. Furthermore, because of the lower melting temperatures of $YAlO_3$ or $Y_3Al_5O_{12}$, key oxide structural evolution processes (e.g., oxide dissolution, nucleation) that occur while $Y_2O_3$ is solid may instead happen while the oxide is molten after reacting with Al.

### 4.5. *Oxide dissolution kinetics*

The degree to which oxide particles dissolve is controlled by the thermal excursion, bulk oxide content, and oxide size. **Fig. 14b** tracks the oxide solubilities for the thermal excursions in **Fig. 14a**. The $Y_2O_3$ solubility along several trajectories far exceeds the maximum $Y_2O_3$ content of ~1–2 at% in conventional ODS alloys. Although such high $Y_2O_3$ concentrations are never attained in the far-field melt, there may be local transient enrichment of $Y_2O_3$ near the oxide-metal interface as the oxides dissolve. The timescale over which the solubility exceeds the bulk $Y_2O_3$ concentration is of order 50 μs, providing a characteristic timescale for dispersoid dissolution.

The dissolution kinetics of $Y_2O_3$ in Ni are not well characterized. However, the dissolution reaction is expected to proceed rapidly at the elevated temperatures in the melt pool, suggesting that the overall oxide dissolution process will be rate-limited by diffusion. Accordingly, we assume that the concentration of dissolved $Y_2O_3$ at the oxide-alloy interface instantaneously reaches the equilibrium value determined by **Eq. 9** and the local temperature. Given a lack of data on temperature-dependent diffusion in liquid Ni [57], the diffusivity of all species is approximated as $D = D_0 \exp\left(-\frac{E_A}{RT}\right)$, with $D_0$ and $E_A$ taken from the diffusion of O in liquid Fe [34]. This simplification is justified since chemical diffusivities in liquid alloys are typically of order $10^{-9}$ m$^2$/s near the alloy melting temperature [58]. $D$ ranges from $1.6 \times 10^{-9}$ to $2.1 \times 10^{-8}$ m$^2$/s between the melting and vaporization temperatures of Ni-20Cr.

To estimate the extent of oxide dissolution, we recognize that the diffusion timescale around a spherical particle, $r^2/D$, is of order 1 μs for 100 nm-scale dispersoids, much faster than the typical thermal excursion of order 100 μs. Accordingly we use a quasi-steady state approximation to describe the concentration field around a dissolving particle. Solving the 1D diffusion equation in spherical coordinates gives [59]

$$\frac{dd}{dt} = -\frac{4V_{m,ox}D(x_{eq}-x_\infty)}{V_m d}, \qquad (12)$$

where $x_{eq}$ and $x_\infty$ are the equilibrium and far-field oxide concentrations, respectively. $x_\infty$ is tracked throughout the simulation via mass conservation as the oxide dissolves into and precipitates from the alloy matrix.

**Fig. 16** shows reduction in oxide diameter post-dissolution for various initial oxide diameters ($d_0$), when exposed to the thermal excursions in **Fig. 14**, following the dissolution law in **Eq. 12**. The degree to which an oxide dissolves is strongly affected by both its initial size and the thermal excursion (i.e., its initial location within the melt track). As an example, the $\tilde{z} = 0.25$ curve, which represents dispersoids traversing the center of the melt track, shows that oxides smaller than 4 μm fully dissolve whereas oxides larger than 6 μm are relatively unaffected. The transition from full dissolution to negligible dissolution occurs over a doubling of the oxide diameter, a relatively narrow size range compared to the range of oxide diameters observed in L-PBF ODS alloys. Thus, for a given thermal excursion, there is a critical initial dispersoid size below which oxide particles fully dissolve. This critical dissolution size ($d_{dis}$) shifts towards smaller diameters for trajectories further from the laser spot, with faster, cooler thermal excursions. In **Fig. 16**, this critical size decreases from 5 µm to 20 nm as $\tilde{z}$ increases from 0.25 to 1.75.

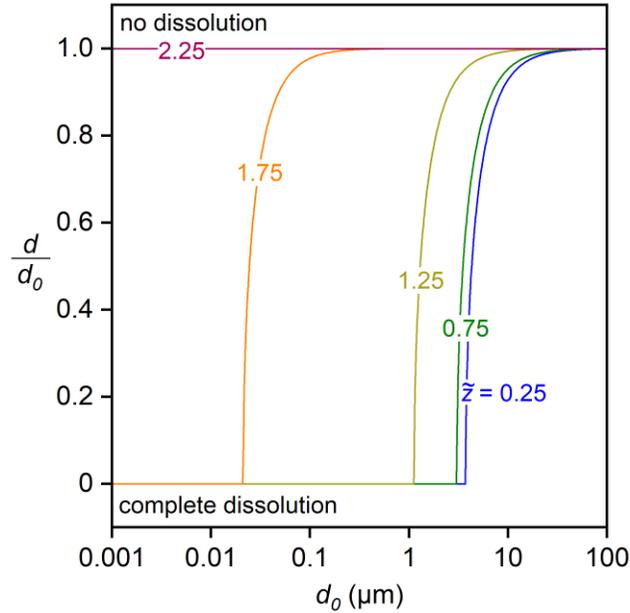

**Fig. 16.** Ratio of dispersoid diameter after dissolution ($d$) to initial oxide diameter ($d_0$) as a function of initial dispersoid size for the thermal excursions shown in **Fig. 14** ($P$ = 140 W, $v$ = 1.2 m/s).

This critical dissolution size has important implications for dispersoid evolution and slag formation. Dispersoids smaller than 20 nm completely dissolve throughout most of the melt pool. Therefore, if all oxides are smaller than 20 nm prior to the melt cycle, none of the initial oxide dispersoid population is retained after a single melt pass, and the size of as-printed oxides is solely be determined by precipitation phenomena upon cooling. However, the $Y_2O_3$ particles in AM ODS feedstock are typically larger than 20 nm [4,6,12,13,60,61]. For example, our $Y_2O_3$ feedstock contains clusters ranging in size from several 100 nm to several µm. Through comminution during RAM, some clusters break down to form a coating of 10-100 nm diameter oxides, but, crucially, micron-scale clusters remain (see **Fig. 1**). Oxides in the size range 100 nm to 1 µm dissolve if they pass through the hot core of the melt pool, but are retained if they pass through the cooler periphery. Oxides that fail to substantially dissolve after one melt cycle may continue to coarsen during subsequent melt cycles, leading to undesirable slag inclusions in the final part. Clearly, $Y_2O_3$ feedstock on the 10 nm-scale is preferred for achieving fine dispersoids. However, larger feedstock may be admissible given judicious selection of printing parameters as discussed in **Sec. 5.1**.

4.6. *Oxide nucleation, growth, and coarsening*

After oxide dissolution during the heating and initial cooling stages of the thermal excursion, the alloy matrix is enriched with dissolved $Y_2O_3$. After the solubility decreases below the equilibrium concentration upon cooling, the dissolved oxide precipitates via homogeneous nucleation [6,14]. Once nuclei have formed, two competing processes consume the remaining supersaturation of $Y_2O_3$: growth and continued nucleation. Growth is defined here as the accretion of dissolved $Y_2O_3$ onto existing oxides over the full size range, consuming material from the supersaturated solution in the far field [62]. Growth is distinct from coarsening, in which larger oxides coarsen at the expense of smaller oxides due to capillary effects [63,64]. Fundamentally, oxide growth and coarsening are significant despite the brief melt cycle because the mean separation of sub-100 nm dispersoids is smaller than the characteristic chemical diffusion distance associated with the duration of the thermal excursion (cf. separation number in **Sec. 4.1**). This section discusses the modeling of nucleation, growth, and coarsening of oxides. These processes proceed concurrently when the alloy is supersaturated with dissolved oxides. For typical L-PBF thermal excursions, the excess $Y_2O_3$ in solution is consumed rapidly through growth, and the $Y_2O_3$ content approaches equilibrium values. Thereafter, oxides may still evolve through coarsening, until $Y_2O_3$ solubility and diffusivity become negligible, typically around the alloy solidification temperature.

Nucleation is modeled with classical nucleation theory, following [14]. The free energy of precipitation, $-\Delta G°(T)$, is the negative of **Eq. 7** in **Sec. 4.4**. Precipitate nuclei with a radius that maximizes the work of formation, i.e., the critical radius, will coarsen monotonically. The critical radius is

$$R^* = \frac{2\gamma_{SL}}{-\Delta G°(T)/V_{m,ox}}, \tag{13}$$

where $\gamma_{SL}$ is the interfacial energy between the oxide and the liquid alloy matrix and $V_{m,ox}$ is the oxide molar volume. The work of formation corresponding to the critical radius is

$$W_R^* = \frac{16\pi\gamma_{SL}^3}{3(\Delta G°(T)/V_{m,ox})^2}. \tag{14}$$

Nucleation rate can then be calculated as follows [62]:

$$I_{ss} = 4\pi Z D(T) R^* \left(\frac{N_A}{V_m}\right)^2 (x_\infty - x_{eq}) x_\infty \exp\left(-\frac{W_R^*}{k_B T}\right), \tag{15}$$

where $N_A$ is Avogadro's constant and $Z$ is the Zeldovich factor. $Z$ is approximated as $\frac{3(\Delta G°(T)/N_A)^2}{4\sqrt{\pi k_B T}(S\lambda_{SL})^{3/2}}$, after [65], where $S$ is a geometric factor equal to $\sqrt[3]{36\pi\Omega^2}$ for spherical nuclei and $\Omega$ is the average atomic volume. $Z$ is of order 0.2 in the present study.

The interfacial energy between the oxide and molten alloy matrix is difficult to estimate and may vary with temperature. If the oxide is pure $Y_2O_3$, then it likely nucleates directly as a solid in molten Ni-20Cr (the present model predicts nucleation occurring over the temperature range 2000-2400K, well below the $Y_2O_3$ melting temperature of 2711K). In this case, the interfacial energy can be estimated from sessile drop studies of liquid Ni-20Cr on a solid $Y_2O_3$ substrate, which give interfacial energies of 1.5–2 J/m² over the temperature range 2000–2400K [23,42].

If $Y_2O_3$ reacted with Al to form $YAlO_3$ or $Y_3Al_5O_{12}$, the oxide may nucleate as a liquid droplet, depending on the temperature at which nucleation sets on (~$T_{dis}$) and is therefore influenced by the oxide concentration in the alloy. As an example, the retained 0.2 at% of $Y_2O_3$ in our experimental samples is expected to nucleate at 2070 K. If the $Y_2O_3$ content is increased to 0.5 at%, nucleation would begin at 2250K, where $YAlO_3$ or $Y_3Al_5O_{12}$ are molten. The interfacial energy between molten oxides and liquid Ni-20Cr is poorly characterized. Further research into interfacial chemistry of molten Y-Al-O phases and the alloy matrix is required for detailed modeling of the oxide evolution when the oxide concentration is in excess of ~0.3 at% in an Al-bearing alloy. As a necessary simplification in this study, the molten oxides are assumed to have the same interfacial energy as solid $Y_2O_3$.

**Fig. 17** is a sensitivity study on the effect of interfacial energy on final dispersoid diameter ($d_f$) and number density (*N*). When the interfacial energy is low, the energetic barrier to nucleation is small, and nucleation proceeds rapidly. Consequently, there are more nuclei on which dissolved $Y_2O_3$ precipitate, leading to a small final dispersoid size and a high dispersoid number density. If the interfacial energy is below 1 J/m$^2$, the as-printed dispersoid diameter may be smaller than 2 nm, much smaller than the dispersoid sizes reported in the literature or observed in the present SANS measurements. The largest final dispersoid size is around 20 nm, assuming an interfacial energy of 2 J/m$^2$. As the interfacial energy increases above 2 J/m$^2$, the alloy must cool to a lower temperature, approaching alloy solidification, for nuclei to form. At such low temperatures, $Y_2O_3$ diffusion becomes too slow for substantial nucleation and growth. For example, with 3 J/m$^2$, dispersoids grow to only 14 nm and have a number density orders of magnitude smaller than with 2 J/m$^2$. In addition, the alloy matrix remains supersaturated with dissolved $Y_2O_3$. Although there are reports of additional dispersoids precipitating during post-process heat treatments [16,29], there is no evidence that the oxides form a homogenized metastable solid solution. Based on this discussion, we set $\gamma_{SL}$ = 2 J/m$^2$ as the sole fitted parameter in the following calculations because it gives predicted final dispersoid sizes that align well with our experimental measurements and is consistent with literature values for the interfacial energy between molten Ni-20Cr and solid $Y_2O_3$ [23,42].

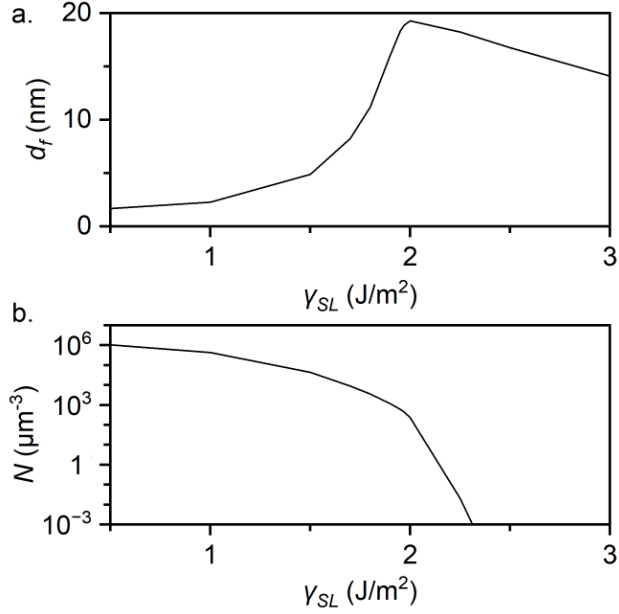

**Fig. 17.** Final dispersoid **(a)** size ($d_f$) and **(b)** number density ($N$) at the center of the melt track shown in **Fig. 13** (P = 140 W, v = 1.2 m/s) as a function of oxide-melt interfacial energy $\gamma_{SL}$.

After dispersoids nucleate, their diameter increases via diffusional growth, according to **Eq. 12**, in the reverse of the dissolution process described in **Sec. 4.5**, and also via coarsening. We model coarsening using classical coarsening theory [63,64]:

$$\frac{\mathrm{d}d}{\mathrm{d}t} = \frac{16 V_{m,ox}^2 \gamma_{SL} D x_\infty}{3 V_m R T d^2}. \tag{16}$$

Coarsening is found to have a minor impact on final dispersoid size because once the dispersoids nucleate, the diffusivity is too slow to support significant coarsening over the brief thermal excursion.

The dispersoid size on cooling was calculated by numerically evaluating **Eqs. 12** and **17** along each particle trajectory, using a forward Euler scheme and accounting for concurrent nucleation, growth, and coarsening. These calculations used the local instantaneous values of temperature, oxide size, chemical diffusivity, and $Y_2O_3$ concentration within the melt at each timestep. Nuclei are assumed to enter the system at the critical radii defined in **Eq. 13**, at a rate determined by **Eq. 15**. Instead of tracking the full dispersoid size distribution, the volume average dispersoid size is computed for the combined population of existing oxides and new nuclei at the end of each timestep. This simplification enables fast calculations required for the dissolution zone model and parameter studies developed in **Sec. 5**. A convergence study was used to select a timestep of 8 ns, corresponding to approximately $10^4$ steps from melting to solidification.

**Fig. 18** tracks the size and number density of a dispersoid population with initial size $d_0$ = 100 nm, following the exemplary trajectories from **Fig. 14**. Trajectories near the center of the melt pool ($\tilde{z} = 0.25, 0.75, 1.25$) exhibit complete dispersoid dissolution. Along these trajectories nucleation sets on when the temperature drops below 2080 K and is complete within ~20 μs.

During and after nucleation, dispersoids simultaneously grow and coarsen. All structural evolution processes halt after the alloy solidifies. The dispersoid size and number density remain constant thereafter.

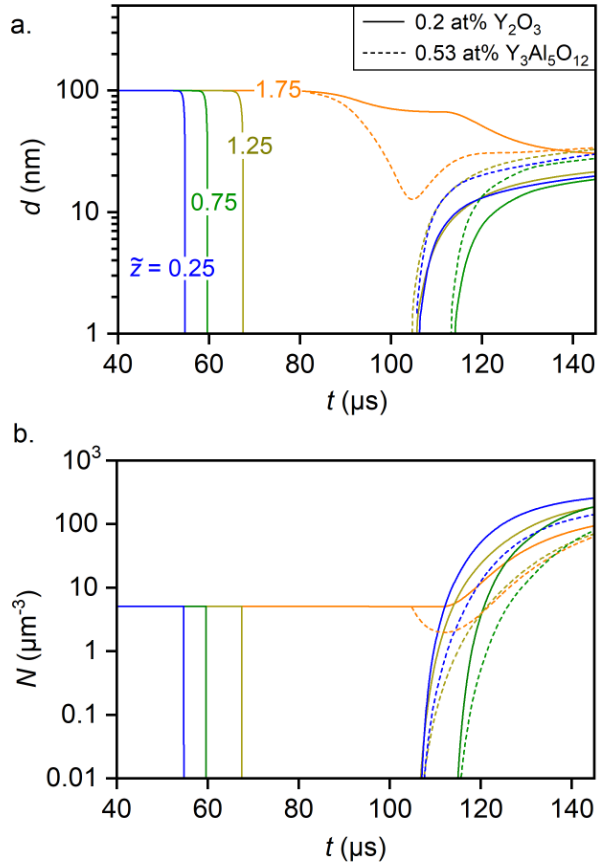

**Fig. 18.** Evolution of dispersoid **(a)** size ($d$) and **(b)** number density ($N$) over the thermal excursions from **Fig. 14**, assuming an initial dispersoid diameter $d_0 = 100$ nm. Solid lines show the evolution of pure $Y_2O_3$ while broken lines show the evolution of $Y_3Al_5O_{12}$, formed through reactions between $Y_2O_3$ and Al while dissolved in the melt.

For trajectories in which dispersoids fully dissolve, there are only minor differences in the final dispersoid size (~20 nm) and dispersoid number density (~100 μm$^{-3}$). This is because all trajectories experience similar cooling profiles in the tail of the melt pool (see **Fig. 14**), thus resulting in similar oxide evolution behaviors after nucleation.

For trajectories near the edge of the melt track (e.g., $\tilde{z} = 1.75$ in **Fig. 18**) 100 nm dispersoids do not fully dissolve. These dispersoids are retained through the hottest portion of the thermal excursion, shrinking to ~70 nm then growing and coarsening upon cooling. Because the timescale for chemical diffusion between the undissolved dispersoids (8 μs) is similar to the dispersoid nucleation timescale (20 μs), the dissolved $Y_2O_3$ is expected to redistribute via two different pathways, plating onto the undissolved dispersoids and also forming new nanoscale dispersoids via nucleation. This results in a bimodal dispersoid size distribution, with one group near the initial size and another group determined by nucleation, growth, and coarsening. The present model only tracks average dispersoid sizes. The full dispersoid size distribution could be modeled using population tracking schemes [43] (cf. [14]); however, detailed population tracking is unnecessary for our central aim of modeling dispersoid evolution because large dispersoids that resist dissolution eventually agglomerate into slag and exit the dispersoid evolution process, an insight used in **Sec. 5.1** to determine conditions for slag formation.

When Al is incorporated into the oxide (see broken lines in **Fig. 18**), the increased oxide solubility in the melt accelerates all diffusion-based structural evolution mechanisms. This leads to faster growth, resulting in a ~60% increase in dispersoid size compared to an Al-free alloy processed under the same conditions. Effects of Al are expanded upon in **Sec. 5.1** to explain slag formation.

4.7. *Dispersoid evolution during single-pass melting*

We now integrate the preceding results into a framework for assessing the oxide structure that results after a single melt pass. Instead of relying on computationally expensive CFD simulations to compute the thermal excursion of each fluid element, we introduce a reduced order model for the steady-state temperature field within the melt pool. The temperature field is calculated in two steps. First, we use the Eagar-Tsai model, which only considers heat conduction, to compute the temperature within a semi-infinite body subjected to a distributed heat flux $I_a$, scanning the free surface with speed $v$ [51]:

$$T(x,y,z) = T_0 + \frac{1}{2\pi\lambda}\int_{-\infty}^{\infty}\int_{-\infty}^{\infty} I_a \frac{\exp\left[\frac{v\rho c_p}{2\lambda}((x-x_1)-|\vec{r}-\vec{r}_1|)\right]}{|\vec{r}-\vec{r}_1|} dy_1 dx_1, \quad (17)$$

where the origin is defined as the laser center. The heat flux for a gaussian beam is

$$I_a(x,y) = \frac{2AP}{\pi r_B^2 \exp\left(-2\frac{x^2+y^2}{r_B^2}\right)}. \quad (18)$$

Next, the results from this calculation are scaled to account for surface tension-driven convection, following the procedure in **Appendix C**. The thermal excursion predicted through this two-step procedure agrees reasonably well with the CFD results (cf. **Fig. C.1** in **Appendix C**), especially during the cooling stages of the melt cycle which ultimately determine dispersoid size and number density. The most significant deviation occurs near the peak of the thermal excursion for trajectories close to the surface, where the scaled Eagar-Tsai solution is higher than the CFD result. This discrepancy does not seriously affect the oxide evolution because all oxides of interest (those initially smaller than 1 μm) are fully dissolved in both cases (cf. **Fig. 16**). We do not account for changes in effective absorptivity with degree of keyholing. Consequently, the framework slightly underpredicts the final dispersoid size at high laser powers and slow scan velocities in **Sec. 5.3.**

We use this reduced order model to compute thermal excursions along different locations across the melt pool cross-section. Dissolution, nucleation, growth, and coarsening are modeled over each thermal excursion to predict the final local dispersoid size. **Fig. 19** summarizes results from these calculations, assuming $P = 140$ W and $v = 1.2$ m/s and varying initial dispersoid size. **Fig. 19** shows that, over the standard range of $Y_2O_3$ feedstock sizes, there is a zone in the center of the melt track in which the oxide feedstock fully dissolves and the final $Y_2O_3$ diameter is independent of feedstock size. Within this dissolution zone, the final $Y_2O_3$ diameter is spatially uniform, varying by less than 5% between the center and the boundary. This homogenous dispersoid size results from the relatively uniform cooling rate. Outside the dissolution zone, dispersoid size increases with distance from the laser, approaching the feedstock size at the melt pool boundary.

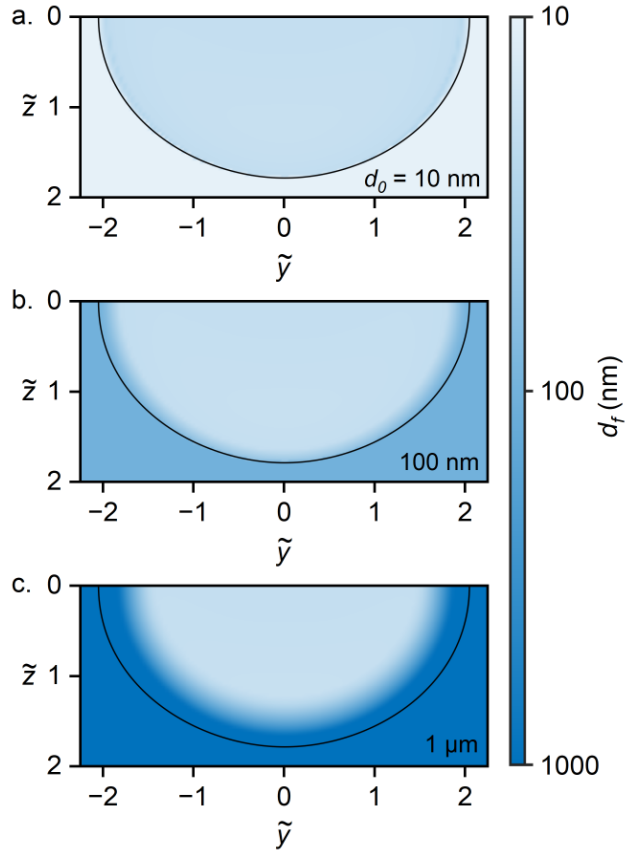

**Fig. 19.** Simulated melt track cross-section for parameter set $P = 140$ W, $v = 1.2$ m/s with colormap for oxide diameter after single-pass melting given initial diameters of **(a)** 10 nm, **(b)** 100 nm, and **(c)** 1 μm.

The present results demonstrate how dissolution can erase details of the initial $Y_2O_3$ feedstock from the final as-printed dispersoid structure. **Fig. 20** presents the same melt track cross-section, overlaid with dissolution zones for varying initial dispersoid sizes. These results show that even 1 μm diameter dispersoids fully dissolve across 60% of the cross-section. With careful control of melt track overlap, as discussed in the following section, a nanoscale oxide dispersion may still be achieved. However, if the feedstock $Y_2O_3$ is much larger than 1 μm, the full dissolution region shrinks to less than half of the melt track, making it impractical to achieve nanoscale dispersoids through dissolution of larger oxide feedstock.

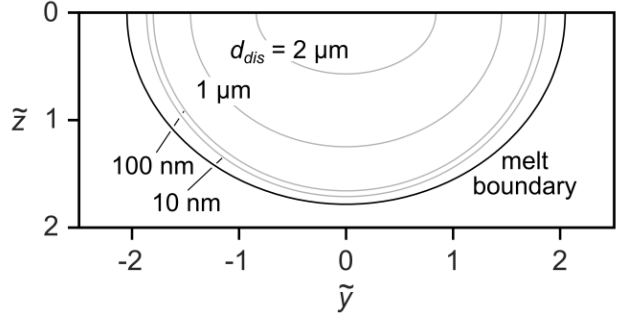

**Fig. 20.** Melt track cross-section for $P = 140$ W, $v = 1.2$ m/s. Dissolution zones for varying oxide diameters are overlaid. Labels indicate the maximum oxide size that will be dissolved within the dissolution zone.

## 5. Implications for L-PBF of ODS alloys

### 5.1. *Dissolution zone model of dispersoid evolution and slag retention*

We have established that the melt pool comprises two regions with disparate dispersoid evolution behaviors: (i) inside the dissolution zone, where oxide particles fully dissolve and subsequently nucleate to form nanoscale dispersoids of spatially-uniform diameter; and (ii) outside the dissolution zone, where oxide particles partially dissolve and the dispersoid sizes range between that within the dissolution zone and the initial size at the melt pool boundary (cf. **Fig. 19**). An important consequence of this finding is that an ensemble of dispersoids which formed in the dissolution zone after a single laser pass will recover to the same diameter during subsequent passes even if it lies outside the dissolution zone. It follows that if the dissolution zones of adjacent melt tracks overlap sufficiently with each other to dissolve large oxides that formed during printing or were present in the powder feedstock, then the dispersoid structure throughout the build volume is homogeneous and matches that from a single pass within the dissolution zone. A key model simplification follows: assuming complete dissolution zone coverage, the dispersoid size in the bulk material is accurately predicted by tracking dispersoid evolution on a single trajectory passing through the center of the melt pool. While this simplification has been implicitly applied in past work [6,14], the present results show that it is conditionally valid through modeling spatial variations in transport-driven dispersoid evolution across the melt pool cross section. This simplification serves as the basis for the dissolution zone model used to predict slag formation and generate process diagrams in what follows.

The volume fraction of printed material that passes through the dissolution zone depends on processing conditions and initial oxide size, and can be assessed using the melt track cross-sections shown in **Fig. 21**. We first define a unit cell with cross-sectional dimensions given by the hatch spacing $\eta$ and layer thickness $\beta$ (blue-shaded rectangle in **Fig. 21**). The build volume can be filled by stacking unit cells, with appropriate interlayer rotation to match the scanning strategy. The volume fraction of dissolution zone material corresponds to the area fraction of the unit cell covered by the dissolution zone (red-shaded area in **Fig. 21**).

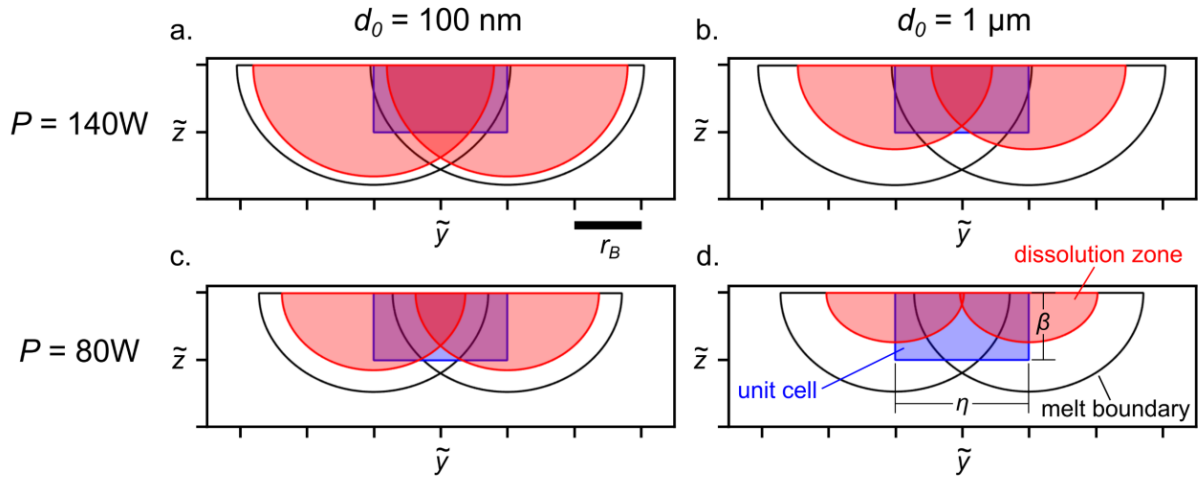

**Fig. 21.** Dissolution zone coverage for different combinations of laser power and initial oxide size. Increasing laser power and decreasing initial oxide size result in greater dissolution zone coverage.

In **Fig. 21a**, dissolution zones from neighboring tracks completely cover the unit cell, resulting in a uniform distribution of ~20 nm dispersoids, formed via nucleation. The dissolution zones in **Fig. 21a** are relatively large, covering most of the melt track, because the initial 100 nm dispersoids are readily dissolved by the thermal excursions throughout most of the melt pool. Increasing the initial dispersoid size to 1 μm as in **Fig. 21b** reduces the dissolution zone coverage. Decreasing the laser power has a similar effect, as shown in **Fig. 21c**. Both of these adjustments decrease the size of the dissolution zone. With hatch spacing and layer thickness kept constant, i.e., unit cell size held fixed, the smaller dissolution zone translates directly to less coverage.

The dissolution zone model reveals how to select printing parameters ($P$, $v$, $\eta$, $\beta$) to ensure adequate dissolution zone overlap, given an initial feedstock $Y_2O_3$ size. This approach is analogous to how melt pool overlap was used in [66] to predict as-printed relative density. Gaps between dissolution zones will result in spatial variations in dispersoid size and multi-modal size distributions, potentially containing oxides large enough for advection-accelerated growth [33]. For example, a feedstock containing micron-scale oxides processed with $P = 80$ W, $v = 1.2$ m/s, $\eta = 40$ μm, and $\beta = 20$ μm will produce a fully dense material but only achieve ~50% dissolution zone coverage (see **Fig. 21d**). Micron-scale oxides outside the dissolution zone will survive the dissolution portion of the thermal excursion and grow larger upon cooling. Micron-scale oxides are large enough to experience relative motion against the surrounding melt (i.e., $Pe > 1$ for oxides larger than 1 μm in **Fig. 11**) and grow at a much faster rate through advection (see **Sec. 4.1**). The scaling analysis in **Fig. 11** further shows that when these oxides grow to ~10 μm, the Stokes number becomes larger than unity. Accordingly, the oxide will deviate from streamlines, impinge upon neighboring oxides, and grow rapidly through agglomeration. In this manner, the combination of the dual adverse conditions—large oxides in the feedstock and poor dissolution zone coverage—may cause runaway oxide coarsening, eventually resulting in slag formation.

## 5.2. *Case studies based on the dissolution zone model*

We next present several case studies to illustrate the predictive ability of the dissolution zone model as well as its limitations. In each study, we assess dissolution zone coverage using the reported values of key experimental parameters (laser power, scan speed, beam radius, hatch spacing, layer thickness, initial oxide size).

Smith et al. used L-PBF to consolidate equiatomic NiCoCr coated with $Y_2O_3$ through RAM [4]. They observed that the as-printed material contained 10-100 μm-scale $Y_2O_3$ inclusions when printed with a large hatch spacing, as shown in **Fig. 22**, whereas the same combination of hatch spacing and other parameters led to defect-free material for the non-ODS variant. In the vertical cross-section shown in **Fig. 22**, the scan direction is alternately angled either 30° or 120° laterally from page-normal. We recreate this scanning condition in the dissolution zone model by laterally stretching both the hatch spacing and dissolution zone by a factor of $1/\cos(\theta)$ to account for the oblique viewing angle $\theta$. The dissolution zone model predicts slag retention zones with periodicity equal to the hatch spacing, in good agreement with experimental observations.

Using the maximum oxide feedstock size of 200 nm reported in [4] as input, the dissolution zones are predicted to cover 98.5% of the bulk volume, as illustrated in the center panel of **Fig. 22**. The remaining 1.5% of the volume, where slag may be retained, is 1 μm tall and clearly too small to accommodate the large slag inclusions observed. The discrepancy may be due to the fact that the oxides are clustered together in the coating prior to L-PBF, similar to what we observed in our RAM-coated Ni-20Cr material. Thus, the dissolution rate is governed by the size of oxide agglomerates rather than the size of individual particles. If the initial oxide size is increased to 1 μm to account for the agglomerated oxide in the feedstock, the slag retention zone grows to 12 μm tall (see right panel of **Fig. 22**), large enough to explain the circular 10 μm-scale inclusions. Furthermore, with an initial oxide size of 1 μm, some slag retention zones in **Fig. 22** are close to those in adjacent layers, enabling slag to grow across multiple layers. This enables the formation of 100 μm-tall inclusions. Smith et al. eliminated the slag issue by decreasing the hatch spacing. Similarly, using this smaller hatch spacing in the dissolution zone model causes the slag retention zones to disappear.

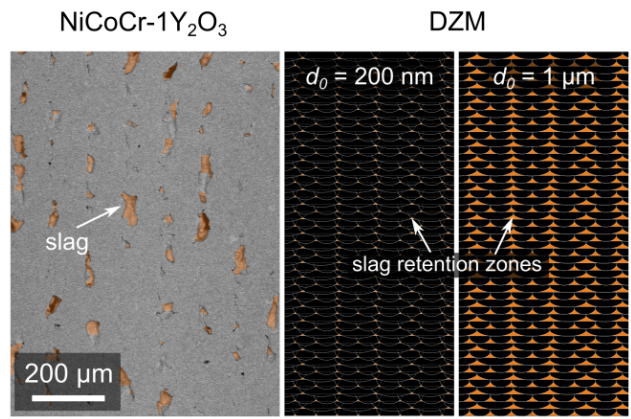

**Fig. 22.** Comparison between SEM observations of slag inclusions in L-PBF NiCoCr printed using the same process as [4] but with increased hatch spacing vs dissolution zone model (DZM) predictions of slag retention zones with $d_0$ = 200 nm or $d_0$ = 1 μm.

Kenel et al. [7] performed L-PBF on Ni-8Cr-5.5Al-1Ti using 20 nm diameter $Y_2O_3$ feedstock incorporated through mechanical alloying. Because of the fine dispersoids in the mechanically alloyed feedstock, the dissolution zone is predicted to cover a larger fraction of the melt track compared to the previous example. However, despite the quality of $Y_2O_3$ feedstock, slag formed during the L-PBF process, as shown in see **Fig. 23**. The explicit process by which nanoscale dispersoids agglomerate into micron-scale inclusions is not considered in the dissolution zone model. However, the dissolution zone model explains how slag is retained after it forms. The large hatch spacing used by Kenel et al. led to only 93% dissolution zone coverage, allowing for slag retention in the remaining 7% of the bulk volume. This is in agreement with the general size and spacing of the slag inclusions observed in experiment. The dissolution zone model can also explain the observations of slag near lack of fusion defects in [7], since the lack of fusion zone will often border the slag retention zone. Thus, parameter sets producing slag defects in ODS alloys will also be susceptible to lack of fusion defects.

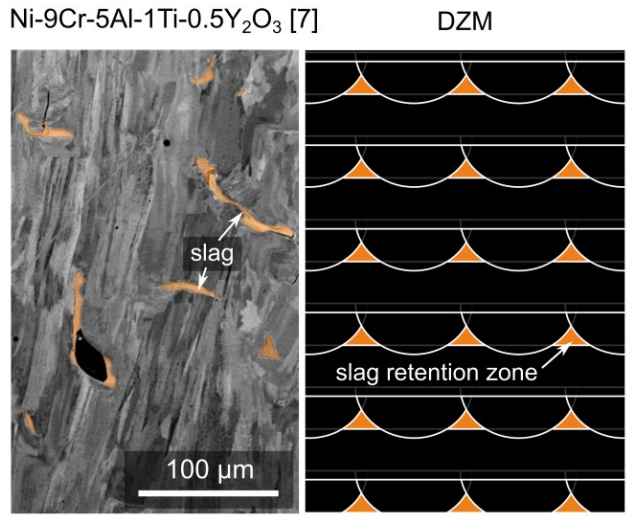

**Fig. 23.** Comparison between SEM observation of slag inclusions in L-PBF Ni-8Cr-5.5Al-1Ti (adapted from [7]) and the dissolution zone model (DZM) prediction.

The dissolution zone model does not explain how the slag extends across multiple layers, as reported by Kenel et al. [7]. We have observed the same phenomenon in our earlier work [10], where it was evident from the eutectic solidification structure in the slag that the slag grew and deformed while molten. Alloying with Al likely exacerbates the slag issue by lowering the melting point of oxides. In **Appendix B**, we show analytically that the oxide melting timescale (<1 μs) is brief compared with the oxide dissolution timescale (100 μs for 1 μm oxides) and residence timescale (~100 μs), establishing that micron-scale oxides will be molten during a substantial portion of the thermal excursion. If the $Y_2O_3$ had reacted with Al, then the oxides would be molten even at the edge of the melt track, where they are deformed by shear flow and grow at accelerated rates. Parts of the molten slag may be transported to the surface and continue to grow in the subsequent layer into the large slag inclusions observed.

In summary, oxide slag in the bulk of as-printed ODS specimens can originate from micron-scale agglomerates of $Y_2O_3$ oxides in the feedstock [4] or from lowering of the oxide melting point by reactions with Al [7,10]. The present framework only considers slag inclusions formed from aggregation of oxides already present in the powder feedstock. The slag inclusions are distinct from surface slag formed through *in situ* oxidation as modeled in [67,68], although the oxygen entrainment models developed in these studies can be integrated with the dissolution zone model. Once micron-scale oxide inclusions are present, they persist and grow unless eliminated through adequate dissolution zone coverage between neighboring melt tracks. This criterion for slag mitigation explains why L-PBF ODS materials require smaller hatch spacings than are typically used with Ni base alloys. From an analysis of the references shown in **Fig. 10**, hatch spacings smaller than the beam diameter typically achieve adequate dissolution zone coverage to suppress slag.

### 5.3. *Processing maps for AM of ODS alloys*

The dispersoid size calculations developed in **Sec. 4.3**-**4.7** were performed over a range of laser powers (60-180 W) and scan speeds (0.6-2.2 m/s), assuming constant beam radius ($r_B = 20$ µm), hatch spacing ($\eta = 40$ µm), layer thickness ($\beta = 20$ µm), and $Y_2O_3$ content (0.16 wt%, consistent with our SANS and WDS measurements). The calculations assumed full dissolution of the initial dispersoids through overlap of dissolution zones. The predicted final dispersoid sizes are plotted as a function of dimensionless printing parameters in **Fig. 24.** There is excellent agreement between the predicted dispersoid sizes and the dispersoid sizes measured through SANS. Deviation at higher powers and lower velocities may result from increased absorptivity due to increased keyholing [32,69], an effect not considered in our thermal model.

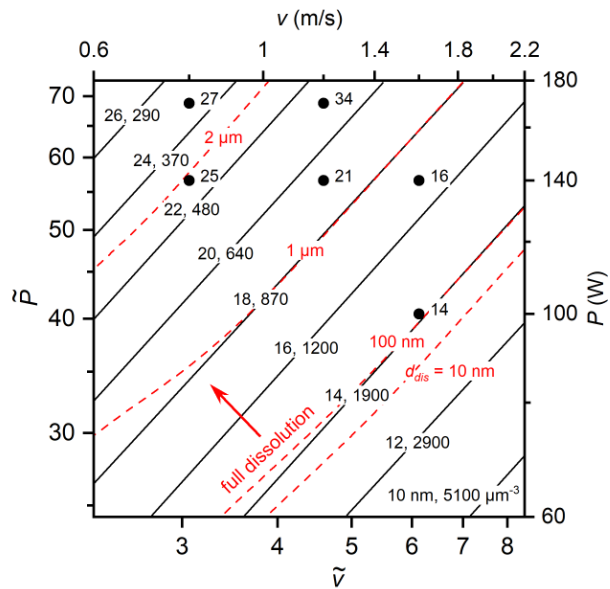

**Fig. 24.** Processing diagram overlaid with black isocontours for predicted final dispersoid diameter and dispersoid number density given a retained $Y_2O_3$ concentration of 0.16 wt%. Experimental parameter sets are labeled with SANS measurements of dispersoid diameter. Red isocontours are bounds for slag-free printing given the initial $Y_2O_3$ size indicated on the contour. Parameter sets towards the upper left are least prone to slag.

**Fig. 24** shows that the dispersoid diameter scales with $\tilde{P}/\tilde{v}$. Increasing the laser power $\tilde{P}$ or decreasing the scan speed $\tilde{v}$ results in slower cooling, allowing more time for dispersoid growth and coarsening and a larger final dispersoid size. The beam radius and materials properties affect dispersoid size through the nondimensional parameters. Dispersoid number densities are indicated alongside corresponding dispersoid sizes in **Fig. 24**. The dispersoid number densities were calculated assuming monodisperse dispersoids, for which

$$N = \frac{6V_{f,ox}}{\pi d^3}. \tag{19}$$

Despite the good agreement in dispersoid size, the number densities predicted by **Eq. 19** are roughly three times larger than those measured through SANS. This discrepancy reflects differences in the dispersoid size distribution; e.g., **Eq. 19** assumes monodisperse oxides in contrast with the log-normal size distribution seen experimentally. Nevertheless, the trends are consistent with experimental measurements, with relatively small increments in dispersoid size translating into large reductions in number density through the $d^3$ dependence in **Eq. 19**.

**Fig. 24** also indicates the lower bound (broken red contours) of processing windows for printing slag-free material for different initial oxide sizes, as predicted by the dissolution zone model. Parameter sets that lie above these bounds have at least 95% of their bulk volume covered by the dissolution zone for the indicated dispersoid size. The bounds approximately follow the final dispersoid size isocontours, although this alignment is coincidental. In the case of slag retention, $\tilde{P}$ and $\tilde{v}$ affect the peak temperature and duration of the thermal excursion, which in turn determine whether oxide particles of a given size can be dissolved. The dissolution zone model bounds predict slag formation in our experimental specimens when the oxide feedstock is larger than ~2 μm, suggesting the slag inclusions shown in **Fig. 3** result from the large $Y_2O_3$ agglomerates that decorate the powder feedstock in **Fig. 1**.

Comparing the slag-free bounds for different initial dispersoid sizes shows that decreasing the oxide feedstock size expands the processing window to lower powers and higher speeds, desirable for high volumetric throughput, high energy efficiency, and fine oxide dispersion. For example, assuming the $Y_2O_3$ feedstock size is 100 nm, the parameter set $\tilde{P} = 40, \tilde{v} = 6$ can produce slag-free material with a dispersoid size of 14 nm and a dispersoid number density of ~2000 μm$^{-3}$. Increasing the initial oxide size increases the minimum achievable dispersoid size and has a similarly deleterious effect on dispersoid number density.

5.4. *Effect of alloy composition on oxide structure*

**Fig. 25** shows the predicted final dispersoid size as a function of $Y_2O_3$ and Al concentration, assuming $P = 140$ W, $v = 1.2$ m/s. Increasing the concentration of $Y_2O_3$ increases the far-field dissolved oxide concentration ($x_\infty$ in **Eq. 12**), which affects dissolution and growth rates. The effect on dissolution rate, which is proportional to $(x_{eq} - x_\infty)$, is limited because $x_\infty$ is typically much smaller than $x_{eq}$ at the high temperatures that drive dissolution. Consequently, the dissolution zone is marginally smaller.

The effect on growth rate is significant because $x_\infty$ is typically much larger than $x_{eq}$ at the low temperatures in which growth takes place. Upon nucleating, dispersoids rapidly grow, consuming the dissolved oxide from the supersaturated melt and suppressing further nucleation. This results in a smaller number of dispersoids which grow to a larger final diameter. The present model predicts that increasing the $Y_2O_3$ concentration from 0.1 to 1 wt% leads to a three-fold increase in dispersoid diameter (15 to 45 nm along the $Y_2O_3$ contour in **Fig. 25**) and a 30% reduction in dispersoid number density. Thus, increasing the bulk $Y_2O_3$ concentration is not a viable strategy for increasing dispersoid number density.

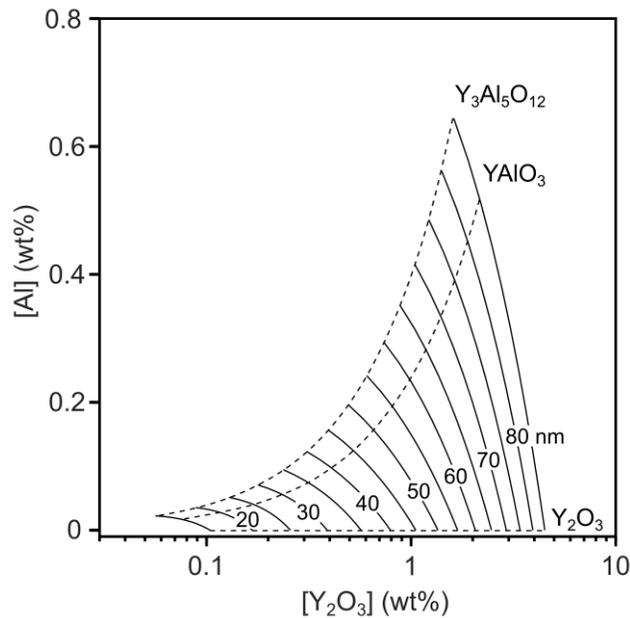

**Fig. 25.** Effects of $Y_2O_3$ and Al concentration on final dispersoid size. Solid lines are isocontours of predicted dispersoid diameter. Broken lines are isocontours of oxide composition, assuming full reaction between Al and $Y_2O_3$.

If Al reacts with $Y_2O_3$ and dissolved $O_2$ to form Y-Al-O phases, the total oxide content increases, and the primary effect is the same as increasing the $Y_2O_3$ concentration. However, there are additional thermochemical effects: Y-Al-O phases have a smaller free energy of formation compared to pure $Y_2O_3$ (114 vs. 123 kJ/gram-atom) This decreases the nucleation rate, resulting in fewer, coarser dispersoids than the same concentration of pure $Y_2O_3$. Following the earlier example, dispersoids predicted to grow to 45 nm when $Y_2O_3$ content is increased to 1 wt% can grow an additional 20 nm when alloyed with just 0.4 wt% Al.

For the retained 0.16 wt% of $Y_2O_3$ in the present experiment, the reaction with Al to form $Y_3Al_5O_{12}$ is predicted to increase dispersoid diameter from 19 to 28 nm. This is consistent with experimental observations of mean as-printed diameter increasing from 21 to 27 nm upon alloying with 1 wt% Al. Similarly, doubling the nominal $Y_2O_3$ concentrations between Ni-20Cr-0.45$Y_2O_3$ and Ni-20Cr-1$Y_2O_3$ (0.16 and 0.36 wt% retained $Y_2O_3$, respectively, assuming similar retention rates) is predicted to increase dispersoid size to 29 nm, in good agreement with the experimentally measured value of 33 nm. Not captured in the present model are effects of Y-Al-O nucleating in molten form, which potentially affects the interfacial chemistry, as described in **Sec. 4.6**. Good agreement between experimental observations and modeling predictions suggest that the effect is minimal over the composition range tested (0.16 wt% $Y_2O_3$, 0-1 wt% Al). Further investigation is required to extend the model to higher Al and $Y_2O_3$ used in commercial Ni-base ODS alloys.

Based on the above analysis, we conclude that Al increases the as-printed dispersoid size by increasing the total dissolvable oxide content and decreasing the nucleation rate. Predictions of the present modeling framework match the experimental observations; there is no obvious indication that Al fundamentally changes the mechanism of oxide evolution for nanoscale dispersoids. In contrast, Al has been linked to slag formation and retention [7,10], suggesting that Al activates alternative growth mechanisms at the larger micron-scale. From these observations, we speculate that fluid mechanics effects and shear deformation become significant in oxide particles growing past the 10 nm-scale. Pure $Y_2O_3$ oxides remain solid and are unaffected, whereas Y-Al-O phases melt and deform under shear, causing runaway growth as described in **Sec. 5.1**.

## 6. Conclusions

Recent demonstrations of L-PBF consolidation of net-shaped ODS alloys have indicated important differences in the oxide structure of L-PBF ODS alloys vs. their wrought counterparts. Specifically, L-PBF ODS alloys have larger dispersoids with lower number densities, and can develop slag inclusions under certain combinations of printing conditions and alloy chemistries. The present work reveals the physics underlying these differences through a combination of systematic L-PBF experiments, scaling analysis, and physics-based modeling. Further, we use these insights to develop the dissolution zone model, which can predict dispersoid structure and slag tendency, providing a tool for co-optimization of alloy chemistry and L-PBF processing conditions to achieve wrought-like structure, properties, and performance.

The present SANS measurements of the full dispersoid size distribution in bulk L-PBF ODS specimens conclusively establish trends in oxide structure with printing conditions and alloy chemistry. We find that the dispersoid size of L-PBF ODS alloys matches that of wrought ODS alloys, while the dispersoid volume fraction and number density are roughly an order of magnitude lower. The latter feature arises from dispersoid loss during printing, potentially because of spatter and slag formation. Our SANS measurements also established that Al alloying above a threshold level, between ~0.34 and 1 at%, promotes reactions between Al and $Y_2O_3$, increases the final dispersoid size, and catalyzes slag formation. Increasing the scan speed and lowering the beam power both decrease the mean dispersoid size, but the processing window and minimum dispersoid size are constrained by the onset of lack-of-fusion defects.

Flow around dispersoids smaller than 100 nm is characterized by extremely small Stokes numbers below $10^{-5}$. Consequently, dispersoids follow streamlines and see no relative motion towards nearby dispersoids or surfaces. Thus, mechanical impingement is unlikely to be significant for dispersoid coarsening in L-PBF. Dispersoids attain their as-printed size primarily through diffusion-mediated growth from the supersaturated melt upon cooling. By contrast, larger micron-scale oxides—either present in the feedstock or coarsened from dispersoids—detach from the flow and experience accelerated coarsening due to mechanical impingement and advective transport of dissolved oxides.

In our modeling we assessed the dispersoid behaviors by coupling a reduced order model of the thermal excursion within the melt pool with classical models of structural evolution processes – e.g., oxide dissolution, nucleation, growth, and coarsening. Our analysis showed that the brief thermal excursion near the core of the melt pool can fully dissolve oxides smaller than ~1 μm. Within this dissolution zone, the final dispersoid size is insensitive to initial location and can be estimated by tracking transport phenomena along a single representative thermal excursion. Upon cooling, nanoscale dispersoids nucleate from the super-saturated melt, explaining how as-printed dispersoids are sometimes smaller than the feedstock oxides. The size and number density of dispersoids produced by nucleation is highly sensitive to the interfacial energy between $Y_2O_3$ and the liquid alloy. Absent alloying additions, pure $Y_2O_3$ is expected to nucleate directly as solid dispersoids with predictable interfacial energy and growth behavior. Mixed Y-Al-O oxides may nucleate as a liquid whose interfacial energy with liquid Ni-20Cr is unknown. The dispersoid size in as-printed material is affected by cooling rate: Fast cooling, achieved through increasing scan speed or decreasing laser power, produces smaller dispersoids and correspondingly greater number density. These model predictions are in excellent agreement with our SANS measurements of dispersoid size and offer guidance for minimizing as-printed dispersoid size in ODS alloys over most of the viable processing envelope.

The dissolution zone model also informs slag-free printing of ODS alloys: combining dissolution zones from neighboring melt pools and comparing with the melt track unit cell indicates when undissolved oxides retained in the outer portion of the melt track may grow further during subsequent laser passes to become slag inclusions. We showed through several case studies on as-printed slag structure in the open literature that the dissolution zone model accurately predicts the periodicity and size of slag inclusions as a function of processing conditions and feedstock oxide size. Further, the dissolution zone model explains how judicious selection of sufficiently high laser power and low scan speed for a given hatch spacing and initial $Y_2O_3$ size can ensure complete dissolution of oxides throughout the bulk material and thus mitigate slag formation. However, this necessarily reduces the viable processing envelope, particularly in the region that produces the smallest dispersoids, presenting a tradeoff between feedstock quality and dispersoid size.

Our model shows that Y-Al-O phases, formed through *in situ* reactions between $Y_2O_3$ and Al, increase oxide solubility, thereby accelerating dispersoid growth and increasing the final dispersoid size. However, the increment in dispersoid size is relatively minor (of order ~50%). This cannot by itself explain how small Al additions catalyze formation of micron-scale slag inclusions observed in the present study and in the open literature [7,10]. A more likely explanation for how Al promotes slag, indicated by the present modeling, is that low melting point Y-Al-O phases nucleate as molten droplets, which are more susceptible to shearing and other behaviors that further accelerate oxide growth.

The present study clearly illustrates that the central challenge of printing ODS alloys is selecting processing conditions that balance minimization of dispersoid size with full densification and slag mitigation. Given the initial $Y_2O_3$ diameter of 100 nm and retained $Y_2O_3$ concentration of 0.16 wt% in the present study, the smallest dispersoids achievable through L-PBF in the slag-free processing envelope is around 20 nm. Further decreasing laser power or increasing scan speed results in finer dispersoids at the expense of print defects (slag formation and lack-of-fusion); e.g., increasing scan speed to 1.6 m/s produced 97.5% dense specimens with dispersoid diameter of 16 nm and number density of 600 $\mu m^{-3}$, still less than a third as many dispersoids as in wrought MA754. Closing the gap in dispersoid number density between L-PBF and wrought ODS alloys requires addressing two issues: (i) mitigating loss of retained oxides due to slag and spatter and (ii) offsetting the increased growth rate resulting from the increased oxide content. Spatter loss may potentially be reduced by using feedstock in which the oxide is bonded to or embedded within powder particles (e.g., through gas atomized reaction synthesis [18] or through mechanical alloying followed by plasma spheroidization [70]). The second issue may be addressed through the addition of alloying elements, such as Ti [71], that decrease the interfacial energy between the alloy matrix and the oxide particle. For example, in our past work on L-PBF MA754 [10], we observed Ti decorating dispersoid surfaces, perhaps modifying interfacial chemistry and growth kinetics. Implementing this approach would complicate the oxide chemistry and requires a deeper understanding of nucleation kinetics and fluid mechanics of molten oxides suspended within the alloy melt.

**CRediT author statement:**

**Wenyuan Hou:** Conceptualization, Investigation, Formal analysis, Writing - Original Draft. **Timothy Stubbs:** Conceptualization, Investigation, Writing - Review & Editing. **Lisa DeBeer-Schmitt:** Investigation, Resources. **Yen-Ting Chang:** Methodology, Investigation. **Marie-Agathe Charpagne:** Supervision, Writing - Review & Editing. **Timothy Smith:** Resources, Writing - Review & Editing. **Aijun Huang:** Conceptualization, Supervision. **Zachary Cordero:** Conceptualization, Supervision, Writing - Review & Editing, Project administration.

**Acknowledgements:** The authors from MIT gratefully acknowledge support from ONR through contract no. N00014-22-1-2036. The authors from Monash University gratefully acknowledge funding from Australia Research Council grant no. LP220100400. A portion of this research used resources at the High Flux Isotope Reactor, a DOE Office of Science User Facility operated by the Oak Ridge National Laboratory. The authors acknowledge the use of instruments and scientific and technical assistance of Dr. Nilanjan Chatterjee at the Department of Earth, Atmospheric, and Planetary Sciences, MIT.

## Appendix A. Approximation formulae for melt pool velocities

The characteristic fluid velocity in the melt pool can be approximated using an idealized unidirectional flow field along the scan direction, a reasonable assumption due to the high aspect ratio of the melt pool. The bottom boundary is stationary from the no-slip condition. The top boundary is subjected to traction from the Marangoni effect:

$$\tau = \frac{d\gamma_{LG}}{dT}\frac{dT}{dx}. \tag{20}$$

The temperature gradient $\frac{dT}{dx}$ is of order $\frac{T_b - T_m}{l}$. The surface tension coefficient $\frac{d\gamma_{LG}}{dT}$ is of order $5 \times 10^{-4}$ N m$^{-1}$ K$^{-1}$ for Ni-20Cr [41]. Accordingly, the velocity gradient $\frac{dv_x}{dz}$ is negative at the surface of the melt pool, and the flow is opposite the scan direction at the surface, as shown in **Fig. A.1a**. After imposing the condition of zero net flow in the scan direction, the flow field can be solved analytically [59]. For the scaling analysis, we are interested in the maximum speed in the melt pool, which represents a worst case for the onset of accelerated coarsening through advection. The maximum speed occurs at the surface and is

$$v_{melt} = \frac{\tau h}{4\mu} = \frac{(T_b - T_m)(d\gamma_{LG}/dT)}{4\mu}\frac{h}{l}. \tag{21}$$

Note that $v_{melt}$ scales with the melt pool aspect ratio $h/l$ and has no other geometric dependence. If we assume DED melt pools have roughly the same aspect ratio as L-PBF, then the speed of convection will be of the same order given the same materials properties (~5 m/s for Ni-20Cr). The scaling analysis is in good agreement with the CFD result in **Fig. A.1b**.

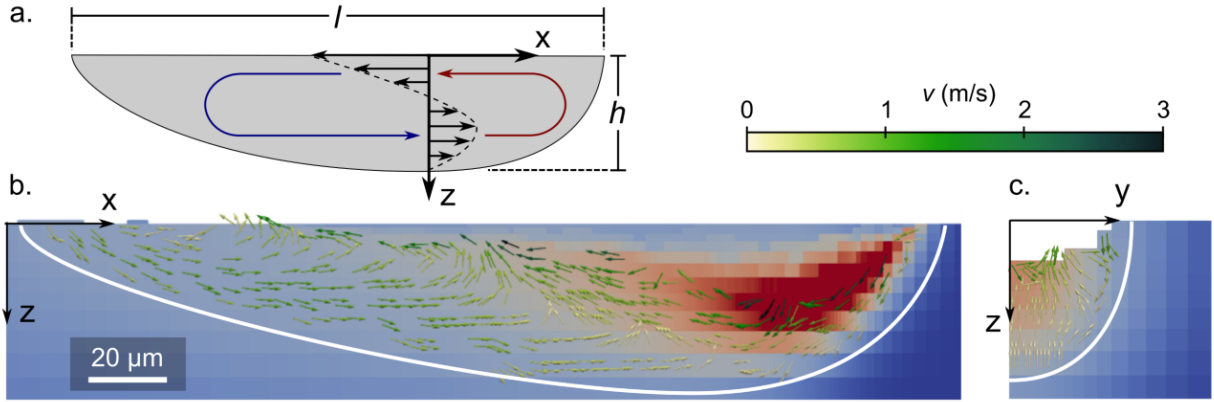

**Fig. A.1. (a)** Schematic velocity field within the melt pool. **(b-c)** CFD calculations of the velocity field, viewed along the symmetry plane and transverse to the scan direction.

To model the effect of melt pool-scale flow on local flow around a dispersoid particle, we introduce a characteristic acceleration of melt pool-scale flow given by

$$a_{melt} = \frac{2v_{melt}}{t_{res}}, \tag{22}$$

i.e., the average acceleration required to bring fluid parcels from stationary to $v_{melt}$ over half the melt pool residence time. A particle suspended in the accelerating fluid will move relative to the fluid due to inertia. For a solid spherical particle, the relative velocity is given by Stokes' law [33]:

$$v_{rel} = \frac{\Delta \rho a_{melt} d^2}{18\mu}, \tag{23}$$

where $\Delta \rho$ is the density difference between the particle and the matrix.

**Appendix B. Timescale of oxide melting**

The timescale for complete melting of a spherical oxide particle can be estimated by considering heat conduction through the molten oxide layer to solid oxide core. Assuming quasi-steady state (i.e., the melt front velocity is slower than heat conduction velocity), the 1-D heat equation in spherical coordinates is [59]

$$\frac{d}{dr}\left(r^2 \frac{dT}{dr}\right) = 0. \tag{24}$$

The boundary conditions are that: the temperature at the melt front ($r = r_m$) is maintained at the oxide melting temperature ($T_{m,ox}$) and the temperature at the surface of the oxide particle ($r = r_0$) is maintained at the instantaneous alloy temperature ($T_\infty$). Solving **Eq. 24** subject to these boundary conditions and applying Fourier's law gives the heat flux at the melt front:

$$q = -\lambda_{ox} \frac{T_\infty - T_{m,ox}}{r_0 - r_m} \frac{r_0}{r_m}. \tag{25}$$

Assuming the incident heat flux at $r = r_m$ is entirely consumed through the latent heat of fusion due to oxide melting $\Delta h_f$, the velocity of the melt front is related to the heat flux through

$$\frac{dr_m}{dt} = \frac{V_{m,ox}}{\Delta h_{f,ox}} q. \tag{26}$$

Combining **Eqs. 25** and **26** and integrating from $r_0$ to 0 gives the oxide melting timescale:

$$t_{melt} = \frac{\Delta h_{f,ox} \rho_{ox} r_0^2}{6\lambda_{ox}(T_\infty - T_{m,ox})}. \tag{27}$$

We can compare the oxide melting timescale to the oxide dissolution timescale, which can be obtained by integrating **Eq. 12** from **Sec. 4.5**:

$$t_{dis} = \frac{V_m r_0^2}{2DV_{m,ox}(x_{eq} - x_\infty)}. \tag{28}$$

The oxide dissolution and melting timescales both scale with $r_0^2$, so their ratio is independent of oxide size:

$$\frac{t_{dis}}{t_{melt}} = \frac{3V_m \lambda_{ox}(T_\infty - T_{m,ox})}{V_{m,ox} D \Delta h_{f,ox} \rho_{ox}(x_{eq} - x_\infty)}. \tag{29}$$

Evaluating **Eqs. 27** and **28** using typical values for $Y_2O_3$ ($\lambda_{ox} = 5$ $Wm^{-1}K^{-1}$, $\Delta h_{f,ox} = 550$ kJ kg$^{-1}$) shows that the melt timescale is shorter than 1 μs for oxides smaller than 1 μm and much faster than the dissolution timescale for all temperatures above melting, establishing that oxides are in the molten state essentially as soon as the surrounding alloy matrix exceeds the oxide melting temperature. This trend is highlighted in **Fig. B.1** which shows that $t_{dis}/t_{melt}$ is much greater than unity above the oxide melting temperature.

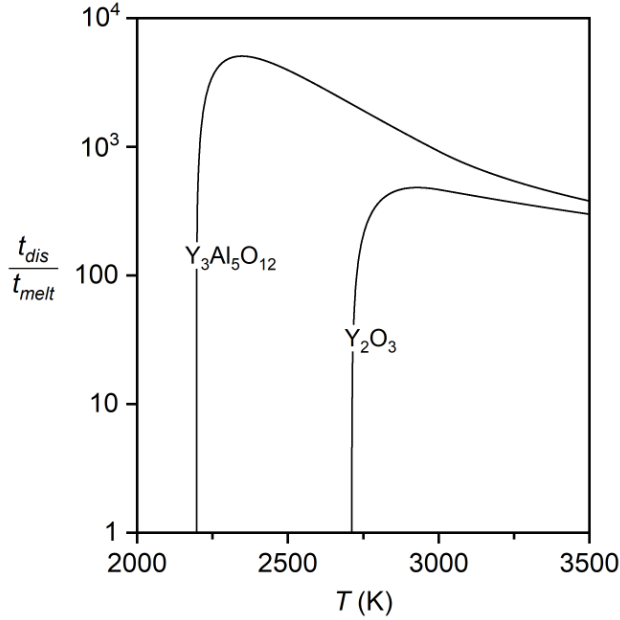

**Fig. B.1.** Ratio of oxide dissolution and melting timescales vs. temperature. $t_{dis}/t_{melt}$ is much greater than unity above the oxide melting temperatures of $Y_2O_3$ (2711 K) and $Y_3Al_5O_{12}$ (2196 K).

## Appendix C. Time-temperature scaling of analytical thermal excursion

We develop a reduced order model for temperature within the melt pool by applying the following empirical transformations to the time-temperature histories predicted by **Eq. 17** in the main text [51]:

$$T_{transformed} = T_0 + B(T - T_0) \tag{30}$$

$$B = \begin{cases} B_1 & T < T_1 \\ B_1 + \left(\frac{T-T_1}{T_2-T_1}\right)^2 (B_2 - B_1) & T_1 \leq T \leq T_2 \\ B_2 & T > T_2 \end{cases} \tag{31}$$

$$dt_{transformed} = dt/C \tag{32}$$

$$C = \begin{cases} C_3 & T < T_3 \\ C_3 + \left(\frac{T-T_3}{T_4-T_3}\right)(C_4 - C_3) & T_3 \leq T \leq T_4 \\ C_4 & T > T_4 \end{cases} \tag{33}$$

**Eqs. 30-31** are applied to temperature while **Eqs. 32-33** transform the time. **Table C.1** lists the best fit parameters for matching the CFD thermal excursions with those predicted through the reduced order model. **Fig. C.1** compares the CFD and the reduced order model predictions showing they are in good agreement, especially during the later stages of cooling.

**Table C.1.** Parameters used in **Eqs. 30-33** to transform the thermal excursion given by **Eq. 17** in the main text.

| Temperature (K) | Scale factor |
|---|---|
| $T_1 = 300$ | $B_1 = 2.0$ |
| $T_2 = 8000$ | $B_2 = 1.3$ |
| $T_3 = 500$ | $C_1 = 1.5$ |
| $T_4 = 8000$ | $C_2 = 6$ |

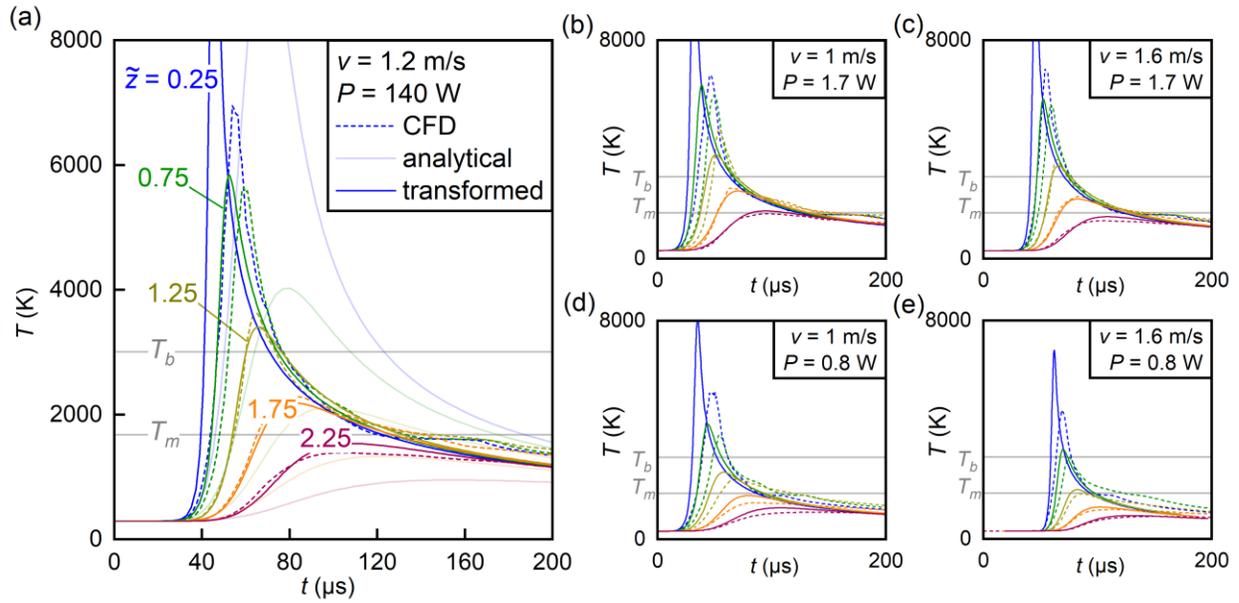

**Fig. C.1.** Comparison between thermal excursions predicted by CFD and by the reduced order model for **(a)** the exemplary parameter set shown in **Fig. 14** and **(b-e)** four parameter sets at the extremes of the viable L-PBF processing envelope. The reduced order model is generally in good agreement with the CFD predictions.